\documentclass{ar-1col-S2O}
\usepackage{url}
\usepackage[numbers]{natbib}
\usepackage{hyperref}
\usepackage{amsmath}
\usepackage{amssymb}
\usepackage{bm}

\jname{Xxxx. Xxx. Xxx. Xxx.}
\jvol{AA}
\jyear{YYYY}
\doi{10.1146/((please add article doi))}

\begin{document}

\markboth{Cano et al.}{Ideal Quantum Geometry for Fractional Chern Insulators}

\title{Ideal Quantum Geometry for Fractional Chern Insulators}

\author{Jennifer Cano$^{1,2}$ and Jie Wang$^{3,4}$
\affil{$^1$Department of Physics and Astronomy, Stony Brook University, Stony Brook, USA, 11794; email: jennifer.cano@stonybrook.edu}
\affil{$^2$Center for Computational Quantum Physics, Flatiron Institute, New York, USA, 10010}
\affil{$^3$International Center for Quantum Materials, Peking University, Beijing, China, 19122; email: jiewang.phy@pku.edu.cn}
\affil{$^4$Beijing Key Laboratory of Quantum Devices, Peking University, Beijing, China, 100871}}

\begin{abstract}
    Quantum geometry plays a fundamental role in many aspects of condensed matter physics. Among its central objects are the Berry curvature and the quantum metric -- quantities that, while distinct, are intertwined through geometric constraints. In this article, we survey recent progress in understanding when and how this bound is saturated, with particular emphasis on the emergence of momentum-space holomorphicity of Bloch states. These developments highlight a profound connection between certain ideal Bloch bands and the Hilbert space structure of the lowest Landau level. We elucidate this relationship through a review of quantum Hall physics in both homogeneous and spatially varying magnetic fields, and conclude by exploring its implications for the search for fractionalized phases in emerging platforms, including moir\'e materials.
\end{abstract}

\maketitle

\tableofcontents

\section{INTRODUCTION} 

Quantum geometry provides a fundamental characterization of the local properties of states in parameter space~\cite{Xiao_2010,Torma_Essay,yu2025quantum,jiang2025revealing,gao2025quantum}.
In condensed matter physics, the quantum geometry of single-particle Bloch wave functions parameterized by crystal momentum captures the fundamental mathematical features governing how Bloch wavefunctions are organized and vary across momentum space. Important implications of quantum geometry in condensed matter physics include the modern theory of electric polarization (Berry connection)~\cite{Vanderbilt_Polarization93,RMP_Resta_Polarization94}, the anomalous velocity (Berry curvature)~\cite{QianNiuSemiClassical99,Haldane_anomalousHall04}, and the localization and spread of Wannier functions (quantum metric)~\cite{RMP_Vanderbilt84}, among other wide-ranging phenomena. The quantum geometry of many-body states also plays an important role, for instance in explaining the quantization of Hall conductivity (many-body Berry curvature)~\cite{Niu_Quantization85}, the optical conductivity sum rules (many-body quantum geometric tensor)~\cite{SWM_SumRule}, and intrinsic and extrinsic dynamics of the fractional quantum Hall effect~\cite{HaldaneGeometryFQH,Du_Graviton,CAN2015752}.

Quantum geometry of Bloch states introduces a characteristic length or energy scale, becoming especially significant in narrow-band systems where electron kinetic energy is quenched. Recently, moir\'e materials have emerged as a class of narrow-band systems in which band dispersion, geometry, and interactions are tunable in situ. In many cases, tuning the material platform, twist angle, and vertical displacement field drives the system into a strongly correlated regime. Various strongly correlated phases of matter have been observed, including superconductivity, correlated insulators, orbital magnetism, and the first observation of the fractional  Chern insulator (FCI)~\cite{andrei2021marvels,nuckolls2024microscopic,Review_moire_FCI}.

The FCI is the lattice analog of the fractional quantum Hall (FQH) effect~\cite{FCI_Bernevig_PRX,sheng2011fractional,neupert2011fractional,parameswaran2013fractional,ZhaoReview}. When it exists in zero magnetic field, it is also referred to as the fractional quantum anomalous Hall effect. Although the zero-field FCI was theoretically predicted in 2011, it was not observed until 2023~\cite{cai2023signatures,zeng2023thermodynamic,park2023observation, xu2023observation,lu2024fractional,xie2025tunable}.

In the search for FCIs in band systems, a variety of factors govern their competition with rival phases such as Fermi liquids and charge-ordered states. These include remote-band effects, band dispersion (electron kinetic energy), and electron–electron interactions. Even in the narrow-band limit, the stability of FCIs remains intricately tied to the interplay between band geometry and interactions.
Further, the notion of ``stability'' itself is multifaceted. It first concerns whether the ground state realizes an FCI or a competing phase.
If an FCI is the ground state, one may then ask how robust it is, e.g., how large is its energy gap and how robust is it against thermal fluctuations~\cite{roy2014band,jackson2015geometric}.
As with other strongly correlated phases, FCI stability is inherently a many-body problem, for which exact criteria are generally elusive except in special solvable limits. Consequently, its study typically relies on a combination of analytical insight and numerical approaches, including exact diagonalization, density matrix renormalization group methods, and diagnostics based on gauge principles and entanglement measures.

From a practical standpoint, a useful -- though not essential -- strategy for realizing fractionalized phases in 2D systems such as moiré materials is to engineer bands that mimic Landau levels, thereby creating favorable conditions to realize FQH-like physics~\cite{parameswaran2012fractional,parameswaran2013fractional,roy2014band,ledwith2020fractional,wang2021exact,mera2021,mera2021b,ledwith2023vortexability,estienne2023ideal}.

In this context, a key question is how to quantify the resemblance of a given narrow band to a Landau level.
The emerging consensus is that band geometry -- particularly the Berry curvature and quantum metric -- play an important role in answering this question. In particular, the so-called trace condition has emerged as a geometric criterion for identifying when a topological band is lowest-Landau-level-like, both in terms of wavefunction structure and the potential to host FCIs as exact ground states. This perspective has proven practically valuable in guiding the search for FCIs in twisted materials.
This article reviews recent development in this direction.

\section{QUANTUM GEOMETRY OF BLOCH WAVE FUNCTIONS}\label{sec:QGT}
\subsection{Geometric properties of Bloch states} \label{sec:geometry}

This review focuses on an isolated single band in two dimensions. Denoting the Bloch wavefunction as $\psi_{\bm k}(\mathbf r) = \langle\mathbf r|\psi_{\bm k}\rangle$, Bloch's theorem prescribes how the wavefunction transforms under lattice translations: $\psi_{\bm k}(\mathbf r+\mathbf a) = e^{i\bm k\cdot\mathbf a} \psi_{\bm k}(\mathbf r)$. Since Bloch wave functions at different momenta have different translation eigenvalues, they are orthogonal; thus, inner products between Bloch wave functions at different momenta vanish. This is not the case for their cell-periodic parts, $u_{\bm k}(\mathbf r) = \langle\mathbf r|u_{\bm k}\rangle = e^{-i\bm k\cdot\mathbf r}\psi_{\bm k}(\mathbf r)$. Thus, the cell-periodic parts of the Bloch wave functions at different momenta can be compared, and this comparison captures the geometric properties of the Bloch band. However, the intrinsic gauge freedom of quantum mechanics, where wavefunctions are defined only up to a phase, renders comparisons subtle.

The Berry connection $\mathbf A_{\bm k} \equiv -i\langle u_{\bm k}|\nabla_{\bm k} u_{\bm k}\rangle$ compares Bloch states by defining parallel transport: given a curve $\gamma$ in the Brillouin zone (BZ), states $|u_{\bm k}\rangle$ with $\bm k \in \gamma$ are said to be parallel if $\mathbf A_{\bm k} = 0$ at all points along this curve. It is always possible to choose a gauge satisfying this condition: breaking the curve into discrete pieces, parallel transport amounts to choosing the phase of $u_{\bm k_{n+1}}$ such that $\langle u_{\bm k_{n+1}}|u_{\bm k_n}\rangle > 0$ is real and positive.

Parallel transport also defines the derivative operator. Since the phase at different momenta is not uniquely defined, only comparisons between parallel states are meaningful. This is accomplished by the covariant derivative operator $D^a = \partial_{k_a} - iA^a_{\bm k}$, which satisfies $\langle u_{\bm k} | D^au_{\bm k}\rangle = 0$.
In the single-band case we consider here, under the gauge transformation $|u_{\bm k}\rangle \rightarrow e^{i\theta_{\bm k}}|u_{\bm k}\rangle$, the derivative operator also transforms as $|D^au_{\bm k}\rangle \rightarrow e^{i\theta_{\bm k}}|D^au_{\bm k}\rangle$.

The crucial subtlety is that along a closed loop $\gamma$, the parallel-transported state need not return with the same phase as the initial state. Such a phase difference between the initial state and the final state after parallel transport along a closed loop is called the Berry phase. The Berry phase is gauge independent, and, by Stokes' theorem, is determined by the integration of Berry curvature density inside the loop. Specifically, if $|u_{\bm{k}_\text{f}}\rangle$ and $|u_{\bm{k}_\text{i}}\rangle$ denote the final and initial states parallel-transported around $\gamma$,
\begin{equation}
    |u_{\bm{k}_\text{f}}\rangle = e^{i \oint_\gamma \mathbf{A}_{\bm k} \cdot d\bm{k} } |u_{\bm{k}_\text{i}}\rangle = e^{i\iint_{M_+} d^2\bm{k} ~ \Omega_{\bm{k}} }|u_{\bm{k}_\text{i}}\rangle = e^{-i\iint_{M_-} d^2\bm{k} ~ \Omega_{\bm{k}}}|u_{\bm{k}_\text{i}}\rangle,
\end{equation}
where $M_{\pm}$ are, respectively, the inner and outer surfaces bounded by the loop $\gamma$ and $\Omega_{\bm k}= \nabla \times \mathbf{A}_{\bm k}$ is the Berry curvature. The choice of $M_+$ or $M_-$ should not matter, which immediately implies the quantization of the Chern number,
\begin{equation}
    C = \frac{1}{2\pi} \int d^2 \bm{k}\,\, \Omega_{\bm k} \in \mathbb{Z},
    \label{eq:Chern}
\end{equation}
which measures the total Berry curvature integrated over the BZ torus.

To summarize: geometrically, the Berry phase characterizes the difference between a state and its parallel-transported counterpart after traversing a closed loop, Berry curvature measures the density of Berry phase, and a non-vanishing Chern number implies the absence of a globally defined smooth gauge choice for the Bloch states.

Having discussed phase, we now turn to the ``distance'' between quantum states. A measure of distance $d(\bm k, \bm k')$ must satisfy three conditions: symmetry $d(\bm k, \bm k') = d(\bm k', \bm k)$; non-negativity $d(\bm k, \bm k') \geq 0$; and the triangle inequality $d(\bm k, \bm k') + d(\bm k', \bm k'') \geq d(\bm k, \bm k'')$. 
The \emph{quantum distance} between Bloch states is defined by $d(\bm k, \bm k') \equiv 1 - |\langle u_{\bm k} | u_{\bm k'}\rangle|^2$, which takes values in $[0, 1]$ (assuming the Bloch state is normalized). Expanding the distance for nearby points gives the \emph{quantum metric},
\begin{equation}
    d(\bm k, \bm k+\delta\bm k) = \sum_{ab} g^{ab}_{\bm k} \delta k_a \delta k_b + \cdots,
    \label{eq:defquantummetric}
\end{equation}
where $a,b = x,y$ denotes the spatial directions.
%

\subsection{Quantum geometric tensor}
The quantum metric, Berry curvature and covariant derivative are nicely organized together in the so-called \emph{quantum geometric tensor} (QGT). In the single band case, the QGT is defined as the inner product of covariant derivative of Bloch states,
\begin{equation}
    Q^{ab}_{\bm k} = \langle D^a_{\bm k} u_{\bm k} | D^b_{\bm k} u_{\bm k} \rangle = g^{ab}_{\bm k} + \frac{i}{2} \epsilon^{ab} \Omega_{\bm k},\label{eq:QGT}
\end{equation}
where $\epsilon^{ab}$ is the anti-symmetric tensor. The QGT is gauge invariant and Hermitian, so that it has a real symmetric part and imaginary anti-symmetric part, which are, respectively, the quantum metric and the Berry curvature. 
The QGT is also positive semi-definite, i.e., for any vector $(c_1,c_2)$,  $c_a^* Q^{ab}_{\bm k} c_b = \langle\chi|\chi\rangle \geq 0$, where $|\chi\rangle = c_a|D^au_{\bm k}\rangle$.

\subsubsection{Coordinate dependence and complex structure} \label{sec:coordinate}
The QGT transforms covariantly under coordinate transformations.
Specifically, under a coordinate transformation $x_a\rightarrow x'_a = \Lambda_{ab} x_b $, with $\det\Lambda = 1$, $\left(Q^{ab}\right)' = \Lambda^{ac}\Lambda^{bd} Q^{cd}$. From Eq.~\eqref{eq:QGT}, $g$ transforms the same way as $Q$, while $\Omega$ is unchanged \cite{parameswaran2013fractional,roy2014band}.

Given a metric $h_{ab}$, with $\det h = 1$, a natural coordinate transformation is given by factorizing it into vectors $\bm e^{1,2}$ (the vielbeins) satisfying
\begin{equation}
    h_{ab} = \delta_{ij} e^i_a e^j_b,\quad \epsilon_{ab} = \epsilon_{ij} e^i_a e^j_b, \label{conditionframe}
\end{equation}
which fixes the vectors' length and orientation.
In the transformed coordinates $\tilde x^{i} = e^i_a x^a$, the ellipse defined by $h_{ab} r^a r^b = R^2$ becomes a circle $\delta_{ij} \tilde x^i \tilde x^j = R^2$.

Such a coordinate transformation also introduces a \emph{complex structure} that will be employed frequently in the subsequent discussion of quantum Hall physics.
A complex structure is a bijective map from the two-dimensional plane $\mathbb R^2$ to the complex plane $\mathbb C^1$; a natural choice is $z = (\tilde x^1 + i\tilde x^2)/\sqrt{2}$.
In the original coordinates, the complex structure is defined concisely by $x \mapsto z = w_ax^a$~\cite{HaldaneGeometryFQH}, where $w_a = (e^1_a + ie^2_a)/\sqrt{2}$.
Eq.~\eqref{conditionframe} then yields:
\begin{equation}
    h_{ab} = w_a^* w_b + w_a w_b^*, \quad i\epsilon_{ab} = w_a^* w_b - w_a w_b^*.
    \label{defcomplexstructure}
\end{equation}
Eq.~\eqref{defcomplexstructure} uniquely fixes the complex structure defined by $h$ up to a global $U(1)$ phase transformation $w_a \rightarrow e^{i\phi} w_a$.
For later use, we introduce the identities
\begin{equation}
    w^aw_a^* = 1, \quad w^aw_a = 0,
    \label{eq:complexstructureidentities}
\end{equation} 
where the indices are raised/lowered by the metric, with $h^{ab} h_{bc} = \delta^a_c$.

\subsubsection{Embedding dependence} \label{sec:embedding}
In tight-binding models with multiple orbitals per unit cell, we refer to changes in the relative positions of orbitals within the unit cell as modifications of the orbital embedding. The QGT is \emph{not} invariant under a change in the orbital embedding \cite{dobardvzic2014effective,jackson2015geometric,simon2020contrasting}.
Generically, a change in embedding adds an extra term to the Berry connection, $\delta \mathbf{A}_{\bm k}$, which leads to a non-vanishing change to both the metric and the Berry curvature. (The latter is given by $\nabla \times \delta \mathbf{A}_{\bm k}$; 
Stokes' theorem ensures that the Chern number in Eq.~\eqref{eq:Chern} is unchanged.)
We return to the importance of the embedding dependence in Sec.~\ref{sec:embeddingmodel}.

\subsubsection{Geometrical bounds}\label{sec:bounds}

The real and imaginary parts of the QGT are not independent. Specifically, the Berry curvature lower-bounds the metric as follows:
\begin{equation}
   \text{Tr}_{h} \, g_{\bm k}  \geq 2\sqrt{\det g_{\bm k}} \geq |\Omega_{\bm k}|,
    \label{eq:inequalities}
\end{equation}
for any positive-definite determinant-one symmetric matrix $h$.
In Eq.~\eqref{eq:inequalities}, we have defined a generalized trace, $\text{Tr}_h g = \sum_{ab} h_{ab} g^{ab}$, which is connected to the basis transformation described in Sec.~\ref{sec:coordinate}: specifically, defining $Q' = \Lambda Q \Lambda^T$, $\text{Tr}\, Q' = \text{Tr}_h Q$ with $h = \Lambda^T \Lambda$.
The proof of the geometric bound in Eq.~\eqref{eq:inequalities} is reviewed in the Supplemental Material (SM) Sec.~\ref{sec:QGTproofs}.
It is equivalent to the integrated version,
\begin{equation}
    \frac{1}{2\pi} \int d^2\bm k ~ \text{Tr}_{h}g_{\bm k} \geq \frac{1}{2\pi} \int d^2\bm k ~ 2\sqrt{\det g_{\bm k}} \geq C. \label{eq:Chernbound}
\end{equation}

\section{QUANTUM HALL PHYSICS}
\subsection{Landau levels in a uniform magnetic field}

In this section, we derive the Landau level wave functions on the plane in a spatially uniform magnetic field, which establishes the notation for our subsequent derivations of the wave functions in a spatially varying field (Sec.~\ref{sec:inhomogeneous}) and on the torus (Sec.~\ref{sec:torus}), as well as of the many-body wave functions (Sec.~\ref{sec:manybodyLLL}).
Much of the material in this section can also be found in several excellent reviews of quantum Hall physics \cite{macdonald1994introduction,girvin1999quantumhalleffectnovel,CAN2015752,tong2016lectures,ArovasQHE}.

The conventional model for quantum Hall physics starts with the effective theory describing the low energy dispersion of a quadratic band,
\begin{equation}
    H_{\rm band} = \frac{1}{2} \left(m^{-1}\right)^{ab} p_a p_b,
    \label{eq:Hamband}
\end{equation}
where $p_a = -i\hbar\partial_a$ is the canonical momentum operator and $m_{ab}$ is the mass tensor. Factoring out its determinant, $|m| = \sqrt{\det m}$, yields a unimodular metric describing its shape,
\begin{equation}
    \left(m^{-1}\right)^{ab} = \frac{1}{|m|} g^{ab},
    \label{eq:defg}
\end{equation}
where $\det g = 1$.

Upon applying an external magnetic field, the canonical momentum operator is replaced by the dynamical momentum operator $\pi_a$ through minimal coupling,
\begin{equation}
    p_a \rightarrow \pi_a(\bm r) = p_a - eA_a(\bm r),
\end{equation}
where $A_a(\bm r)$ is the vector gauge potential whose curl is the magnetic field. Throughout this work, we specialize to two dimensions (2D) and consider a magnetic field normal to the 2D plane so that $\bm B(\bm r) = B(\bm r) \hat{z} $ and $\hat{z}\cdot[\nabla\times\bm A(\bm r)] = B(\bm r)$.

In this section, we consider a uniform magnetic field, so that $B(\bm r) = B$. We choose a convention where $B<0$ and electric charge $e < 0$, so that the combination $eB > 0$. The magnetic length is defined by $\ell_B =\sqrt{\hbar/(eB)}$. The momentum operators have a nontrivial but simple commutator,
\begin{equation}
    \left[ \pi_a, \pi_b \right] = i \hbar^2 \epsilon_{ab}/\ell_B^2.
    \label{eq:picommutator}
\end{equation}

The uniform field model can be solved exactly by transforming the dynamical momentum operators $(\pi_x, \pi_y)$ into a set of canonical ladder operators $(\hat a_h, \hat a^\dag_h)$. Since the ladder operators will be determined by a metric, we add the subscript $h$. Following the notation of the complex structure in Eq.~\eqref{defcomplexstructure}, we define,
\begin{equation}
    \hat a_h = \frac{\ell_B}{\hbar} w^a(h) \pi_a,\quad \hat a^\dagger_h = \frac{\ell_B}{\hbar} [w^a(h)]^* \pi_a,\label{eq:a}
\end{equation}
which satisfy $[\hat a_h, \hat a^\dag_h] = 1$. 
Different choices of $h$ amount to performing a Bogoliubov transformation on the ladder operators~\cite{HaldaneGeometryFQH}.

Rewriting the Hamiltonian in Eq.~\eqref{eq:Hamband} in terms of $\hat a_h$ and $\hat a^\dag_h$ generically contains linear combinations of $\hat a^2_h$, $\hat a^{\dag2}_h$, $\hat a^\dag_h \hat a_h$ and $\hat a_h \hat a^\dag_h$. The model is brought into diagonal form (i.e., without $\hat a^2_h$ and $\hat a^{\dag2}_h$ terms) only for the specific choice $h^{ab}=g^{ab}$. In that case, the Hamiltonian reads,
\begin{equation}
    H = \hbar \omega_c \left( \hat a^\dagger_g \hat a_g + \frac{1}{2} \right) + \frac{1}{2} g \mu_B B,
    \label{eq:HamQHdiag}
\end{equation}
where $\omega_c = eB/|m|$ is the cyclotron frequency and we have added the second term to incorporate the Zeeman splitting, with $g$ indicating the $g$-factor (the distinction with the metric $g$ should be clear from context). For notational simplicity, we now drop the subscript $g$ on the ladder operators.

Eq.~\eqref{eq:HamQHdiag} immediately yields the Landau level energies and eigenvectors.  We now focus on the ``$g=2$'' case, defined by $g=2m_e/|m|$, with $m_e$ the electron mass. In this case,
the last two terms cancel (recall our convention is $e,B<0$, and, as always, $\mu_B = |e|\hbar/2m_e$); 
thus the energy of the $n^\text{th}$ Landau level is $E_n = n\hbar\omega_c$.
The eigenvectors are constructed in the usual way: the $n=0$ states are called the lowest Landau level (LLL), denoted by $|0\rangle$, and are annihilated by $\hat a$. The $n^{\rm th}$ Landau level state is constructed from the lowest one by
\begin{equation}
    |n\rangle = \frac{(\hat a^\dag)^n}{\sqrt{n!}}|0\rangle.
\end{equation}

The LLL wave functions can be found explicitly by working in the symmetric gauge where $A_a(\mathbf{r}) = -B\epsilon_{ab}r^b/2$, and using the identities in Eq.~\eqref{eq:complexstructureidentities}
to write Eq.~\eqref{eq:a} as
\begin{equation}
    \hat a = -i\left(\ell_B \partial_{\bar{z}} + \frac{z}{2\ell_B}\right), \quad \hat{a}^\dagger = i\left(-\ell_B \partial_{z} + \frac{\bar{z}}{2\ell_B}\right)
    \label{eq:aoperator}
\end{equation}
where the complex coordinate $z$ is defined by $z = w_a(g) r^a$, 
$\partial_{\bar{z}} = w^a(g)\partial_a$, and $\bar{z} \equiv z^*$.
It follows that the wave functions annihilated by $\hat{a}$ are given by a holomorphic function (i.e., a function only of $z$, not $\bar{z}$), multiplied by a Gaussian, 
\begin{equation}
    \psi_{\rm LLL}(\bm r) = f(z) \exp\left(-\frac12|z|^2/l_B^2\right).
    \label{eq:psiLLL}
\end{equation}

Inside each Landau level, the states are highly degenerate. Their degeneracy can be resolved by symmetries, as we now explain. The quantum mechanical analog of the classical cyclotron radius is
$\bar{R}^a = -\epsilon^{ab}\pi_b \ell_B^2/\hbar$, which defines a guiding center operator
\begin{equation}
    R^a = r^a - \bar{R}^a = r^a + \epsilon^{ab}\pi_b \ell_B^2 /\hbar.
\end{equation}
From the commutator of dynamical momentum in Eq.~\eqref{eq:picommutator}, the guiding centers obey $[R^a, R^b] = -i\ell_B^2\epsilon^{ab}$. Analogous to the Landau orbital ladder operators $\hat{a}_h, \hat{a}^\dagger_h$, the guiding center operators combined with a choice of metric define guiding center ladder operators,
\begin{equation}
    \hat b_h = [w_{a}(h)]^* R^a/\ell_B, \quad \hat b^\dag_h = w_{a}(h) R^a /\ell_B,
\end{equation}
which satisfy $[\hat b_h, \hat b^\dag_h] = 1$. The differential operators are given explicitly by
\begin{equation}
    \hat{b} = \ell_B \partial_z  + \frac{\bar{z}}{2\ell_B}, \quad \hat{b}^\dagger = -\ell_B \partial_{\bar{z}}  + \frac{z}{2\ell_B},
\end{equation}
where again the complex coordinate $z$ is defined from the complex structure associated to the unimodular metric $h$.

Since the guiding center ladder operators commute with the Hamiltonian, they can resolve the degeneracy of each Landau level. 
The symmetry of the manifold determines how to do so.
On the infinite plane or the sphere, which have continuous rotation symmetry, the degeneracy is resolved by rotational angular momentum $m$, while on the cylinder or torus, which have translation symmetry, the momentum $\bm k$ resolves the degeneracy.

\subsection{Lowest Landau level in an inhomogeneous magnetic field}
\label{sec:inhomogeneous}
In a homogeneous magnetic field, the Zeeman term shifts the energy levels uniformly. If the $g$-factor happens to be $g=2$, then the LLL sits exactly at zero energy. 
We now prove a rather surprising fact: even in a spatially varying magnetic field, at $g=2$, the LLL remains flat and pinned exactly to zero energy~\cite{aharonov1979ground,CAN2015752}.

We split the inhomogeneous magnetic field, $\mathbf{B}(\mathbf r)$, into its uniform part, $\mathbf{B}$, and its spatially varying part, $\delta\mathbf{B}(\mathbf r)$, which averages to zero. We use $A_a(\mathbf r)$ to denote the vector potential associated with the uniform part of the magnetic field, which, without loss of generality, can be written in symmetric gauge as $A_a(\mathbf r) = -B\epsilon_{ab}r^b/2$, and use $\delta A_a(\mathbf r)$ to denote the vector potential associated with the spatially varying part of the magnetic field. Since $\delta\mathbf{B}(\mathbf{r})$ integrates to zero, $\delta \mathbf{B}$ and $\delta \mathbf{A}$ can be written in terms of a scalar field $\phi$ as
\begin{equation}
	\delta A_a(\mathbf{r}) = - \epsilon_{ab} g^{bc}\partial_c \phi(\mathbf{r}),
    \quad 
    \delta B(\mathbf{r}) = g^{ab} \partial_a \partial_b \phi(\mathbf{r}) \equiv \Delta_{\mathbf r} \phi(\mathbf{r}), \label{eq:laplacianphi}
\end{equation}
where $\Delta_{\mathbf r} = g^{ab} \partial_a \partial_b$ is the Laplacian operator. 
When the full magnetic field can be written as the Laplacian of a scalar potential, i.e., $\mathbf B(\mathbf r) = \frac{\hbar}{e} \Delta_{\mathbf r} K(\mathbf r)$, we refer to $K(\mathbf{r})$ as the K\"ahler potential.
In the present case,
\begin{equation}
    K(\mathbf r) = \frac{1}{4l_B^2} g_{ab} r^a r^b + \frac{e}{\hbar} \phi(\mathbf r) = \frac{1}{2l_B^2} |z|^2 + \frac{e}{\hbar} \phi(\mathbf r). \label{defKahlerpotential}
\end{equation}


Denoting the dynamical momentum of the inhomogeneous field as $\pi_a(\mathbf r)$ and using $\pi_a$ to denote the momentum in the uniform magnetic field,
\begin{equation}
    \pi_a(\mathbf{r}) = \pi_a - e\delta A_a(\mathbf{r}) = \pi_a + e \epsilon_{ab} g^{bc} \partial_c \phi(\mathbf r),
\end{equation}
whose commutator now depends on the total magnetic field,
\begin{equation}
   ~[\pi_a(\bm r), \pi_b(\bm r)] = ie\hbar \epsilon_{ab} B(\bm r).\label{alg_pi}
\end{equation}

We now consider the $g=2$ problem in such an inhomogeneous magnetic field, described by the Hamiltonian
\begin{equation}
    H = \frac{1}{2m} g^{ab} \pi_a(\bm r) \pi_b(\bm r) + \frac{1}{2} g \mu_B B(\bm r).\label{eq:HamQH}
\end{equation}
The spatial dependence of the commutator in Eq.~(\ref{alg_pi}) makes the model in Eq.~(\ref{eq:HamQH}) impossible to solve analytically. 
Despite this difficulty, we will show that the zero mode remains, i.e., the LLL remains completely degenerate.

To start, we use the complex structure $w_a$ associated with the mass tensor (i.e., the complex structure defined in Eq.~\eqref{defcomplexstructure} with the metric $g_{ab}$ given in Eq.~\eqref{eq:defg}) to introduce the holomorphic and anti-holomorphic momentum operators,
\begin{equation}
    \pi^{-}(\bm r) = w^a\pi_a(\bm r),\quad \pi^{+}(\bm r) = (w^a)^* \pi_a(\bm r),
\end{equation}
which are analogous to the operators in Eq.~\eqref{eq:a}, but have a spatially varying commutator,
\begin{equation}
    ~[\pi^{-}(\bm r), \pi^{+}(\bm r)] = e\hbar B(\mathbf{r}).
\end{equation}
Using these operators, the Hamiltonian in Eq.~\eqref{eq:HamQH} can be rewritten as:
\begin{equation}
    H = \frac{1}{m} {\pi}^{+}(\mathbf{r}) {\pi}^{-}(\mathbf{r}).
    \label{eq:HamQHinhomogenous}
\end{equation}
Thus, the kernel of $\pi^-(\mathbf r)$ is precisely the zero energy eigenspace of $H$. 
The dimension of this space is unchanged by the varying magnetic field because the $\pi^-(\mathbf r)$ operator is related to its counterpart in the uniform field, $\pi^- = \hat{a}\hbar/\ell_B =  -i\hbar\bar\partial - i\hbar z/(2l_B^2)$, by the similarity transform
\begin{equation}
    \pi^{-}(\bm r) = w^a\pi_a(\bm r) = \pi^- -ie\bar\partial\phi(\mathbf r) = e^{-\frac{e}{\hbar}\phi(\mathbf r)} \, \pi^- \, e^{\frac{e}{\hbar}\phi(\mathbf r)}.
    \label{eq:similaritypi}
\end{equation}
Hence, if $|0\rangle$ denotes the state annihilated by $\pi^-$ in a homogeneous magnetic field, then $e^{-e\phi(\mathbf{r})/\hbar}|0\rangle$ will be annihilated by $\pi^-(\mathbf{r})$ in an inhomogenous field, and thus is a zero energy eigenstate of the Hamiltonian in Eq.~\eqref{eq:HamQHinhomogenous}.
It follows that the entire LLL remains exactly at zero energy in an inhomogeneous field.
In terms of the K\"ahler potential introduced in Eq.~\eqref{defKahlerpotential}, Eq.~\eqref{eq:similaritypi} also implies that the LLL wave function in Eq.~\eqref{eq:psiLLL} can be written as
\begin{equation}
    \psi_\text{LLL} (\mathbf{r}) = f(z) e^{-K(\mathbf{r})},
    \label{eq:psiLLLkahler}
\end{equation}
which applies to both the homogeneous or inhomogeneous magnetic field cases.
We note that the degenerate LLL is a unique feature of the zeroth Landau level: in an inhomogeneous field, the higher Landau levels are not flat in energy.

\subsubsection{Dirac Landau levels in magnetic field}
We now briefly compare the charged particle in a magnetic field to the 2D Dirac fermion in a magnetic field, which is closely related to the $g=2$ quantum Hall problem just described. We will show that it also has a flat LLL that survives spatial variation in the magnetic field. 
Physically, the model can be realized at the interface between a Type II superconductor and a Dirac material such as graphene or a topological insulator surface~\cite{JieLiangFuDirac22}.

A 2D Dirac fermion in a non-uniform magnetic field is described by the Hamiltonian:
\begin{equation}
    H_{\rm Dirac} = v_F\sum_{i=1,2} \sum_{a=x,y}\frac{1}{\sqrt{2}} \sigma^i e_i^a \pi_a(\bm r) =   v_F \left(\begin{matrix} & \pi^+(\bm r) \\ \pi^-(\bm r) & \end{matrix}\right),
\end{equation}
where $e_i^a$ is the frame field introduced in Sec.~\ref{sec:coordinate}, which accommodates an anisotropic Dirac cone. When the magnetic field is uniform, the above model reduces to the standard problem, which exhibits a zero mode and a particle-hole symmetric energy spectrum, $E_n = \pm \hbar v_F/\ell_B \sqrt{n}$, describing the famous Landau level spectrum of graphene \cite{novoselov2005two,zhang2005experimental,gusynin2005unconventional,peres2006electronic}. 
When the magnetic field is spatially modulated, the full spectrum cannot be found analytically. However, the zero mode can be found by squaring the Hamiltonian:
\begin{equation}
    H_{\rm Dirac}^2 = v_F^2 \left( \begin{matrix} \pi^+(\bm r)\pi^-(\bm r) & 0 \\ 0 & \pi^-(\bm r)\pi^+(\bm r) \end{matrix} \right) 
\end{equation}
The upper quadrant is precisely the $g=2$ quantum Hall problem in Eq.~\ref{eq:HamQHinhomogenous}. Hence, the zero mode survives the spatially modulated field, and its wave functions are given by
\begin{equation}
    \tilde\Psi_0 = \left(\begin{matrix} f(z) \\ 0 \end{matrix}\right) e^{-K(\bm r)},
    \label{eq:zeroDirac}
\end{equation}
where $K(\mathbf r)$ is the K\"ahler potential in Eq.~\eqref{defKahlerpotential}.
Eq.~\eqref{eq:zeroDirac} is a spinor-generalization of Eq.~\eqref{eq:psiLLLkahler}.

\subsection{Landau level wave functions on the torus} \label{sec:torus}
The quantum Hall problem on the torus describes electrons on a 2D manifold with periodic boundary conditions in both directions and a uniform perpendicular magnetic field. Since the torus is a special case of the infinite plane with boundaries identified, the LLL states on the torus must also be holomorphic functions, locally identical to Eq.~\eqref{eq:psiLLL}. However, the periodicity strongly constrains the analytical form of the global wave function~\cite{haldanetorus1,haldanetorus2}.

\subsubsection{Boundary conditions and translation operators} \label{sec:torusbc}
First, the torus geometry requires the magnetic flux quanta be an integer, $N_\phi$, due to the Dirac quantization condition. Because each flux corresponds to an area of $2\pi l_B^2$, the torus area must be,
\begin{equation}
    \hat{z}\cdot\left(\bm L_1 \times \bm L_2\right) = 2\pi N_\phi l_B^2,
    \label{eq:torusflux}
\end{equation}
where $\bm L_1$ and $\bm L_2$ are respectively the two primitive real space vectors spanning the torus and $\hat{z}$ is the unit vector normal to the torus surface. We orient the vectors such that $\bm L_1 \times \bm L_2$ is positive.
To fully specify the quantum Hall problem on the torus requires, in addition to the primitive vectors, boundary conditions,
\begin{equation}
    t(\bm L_i) |\psi\rangle = e^{i\theta_i} (-1)^{N_\phi} |\psi\rangle,\label{def:bc}
\end{equation}
where $t(\bm L_i)$ is the translation operator that transports an electron around the torus and $0 \leq \theta_{i=1,2}<2\pi$. The boundary conditions can be interpreted as threading fictitious flux through a hole of the torus, so that $\theta_i$ is simply the Berry phase acquired by an electron transported around the torus cycle.

To understand the constraint of the boundary condition in Eq.~\eqref{def:bc} requires the action of translation operators in the LLL. 
The magnetic translation operator that transports an electron by a distance $\bm d$ is given by,
\begin{equation}
    t(\bm d) = \exp(i\bm d\times\bm R/\ell_B^2) = \exp\left[(d^*\hat{b}^\dag - d \hat{b})/\ell_B\right],
\end{equation}
where $d=w_ad^a$ is the complex coordinate of $\bm d$. 
In the LLL, the translation operator acts by (see SM Sec.~\ref{sec:translationLLL}),
\begin{equation}
    t(\mathbf{d}) \psi_{\text{LLL}}(\mathbf{r}) = e^{\frac{1}{2}\left( z\bar{d} - \bar{z} d \right) /\ell_B^2} \psi_{\text{LLL}}(\mathbf{r} - \mathbf{d}), \label{eq:translationLLL}
\end{equation}
where $\psi_{\text{LLL}}(\mathbf r) = f(z) \exp(-|z|^2/2l_B^2)$. 
Thus, the translation operators do not commute, and their non-commutativity is determined by the Berry phase accumulated by an electron traversing the closed loop spanned by $\bm d$ and $\bm d'$:
\begin{equation}
    t(\bm d) t(\bm d') = e^{i\bm d\times\bm d'/l_B^2} t(\bm d') t(\bm d).
\end{equation}
As mentioned below Eq.~\eqref{def:bc}, physically $t(\bm d)$ can be achieved by fictitious flux insertion. Since, by gauge invariance, the Hilbert space is only invariant under insertion of integer flux, the allowed translations must be discrete and commute with $t(\bm L_i)$. Thus, the elementary magnetic translation group elements are given by $\hat t_{i} = t(\bm L_i/N_\phi)$; their commutation relation $\hat t_1 \hat t_2 = \exp(2\pi i/N_\phi) \hat t_2 \hat t_1$ implies an $N_\phi$ degeneracy of each Landau level.

Combining the form of the wave function in Eq.~\eqref{eq:psiLLL} with the boundary condition in Eq.~\eqref{def:bc}, and applying the action of the translation operator in Eq.~\eqref{eq:translationLLL} yields 
\begin{equation}
    f(z + L_i) = e^{-i\theta_i} e^{\varphi_i(z)} (-1)^{N_\phi} f(z), \label{defbclll}
\end{equation}
where $L_i$ is the complex coordinate of the primitive torus vectors and $\varphi_i(z)=L_i^*(z + L_i/2)/\ell_B^2$, which 
is holomorphic in $z$, as required.

\subsubsection{Lowest Landau level wavefunctions on the torus}
\label{sec:LLLwftorus}
The boundary condition in Eq.~\ref{defbclll} combined with the expression for the torus area in Eq.~\eqref{eq:torusflux} can be used to evaluate the following contour integral around the torus~\cite{haldanetorus1,haldanetorus2},
\begin{equation}
    \frac{1}{2\pi i} \oint dz ~ \partial_z \ln \left[f(z)\right] 
    = \frac{1}{2\pi i}\left[ \int_0^{L_1} dz~ \partial_z \ln \frac{f(z)}{f(L_2 + z)} + \int_0^{L_2} dz~ \partial_z \ln \frac{f(L_1+z)}{f(z)} \right] 
    =  N_\phi.
    \label{eq:Nphizeros}
\end{equation}
This implies that each LLL wavefunction on the torus must have $N_\phi$ zeros. 
The periodicity and locations of the zeros uniquely fix the analytical form of the LLL wavefunction on the torus: for zeros at positions $\omega_1, \dots, \omega_{N_\phi}$, the wave function can be written as~\cite{haldanemodularinv,JW_MonteCarlo},
\begin{equation}
    \Phi_{\bm\omega}(\bm r) = \left[\prod_{i=1}^{N_\phi} e^{\omega_i^* z/(N_\phi\ell_B)} \tilde\sigma(z - \omega_i)\right] e^{-\frac12|z|^2/l_B^2},\label{defsingleparticlewf}
\end{equation}
where $\tilde{\sigma}(z)$ denotes the modified Weierstrass sigma function~\cite{haldane2018modular} (a holomorphic function containing one zero in the fundamental domain) with the following quasi-periodicity
\begin{equation}
    \tilde\sigma(z + L_i) = - e^{L_i^*\left( z + \frac{L_i}{2} \right) / (N_\phi\ell_B^2) } \tilde\sigma(z).
    \label{eq:quasiperiodicitytorus}
\end{equation}
The Weierstrass sigma function is used instead of the Jacobi theta function because it has more compact translation properties and is modular invariant.

However, the positions of the zeros cannot be chosen arbitrarily: their sum is constrained by the boundary conditions on the torus to satisfy
\begin{equation}
    \Box \equiv \sum_{n=1}^{N_\phi} \omega_n = \frac{\theta_1}{2\pi} L_2 - \frac{\theta_2}{2\pi} L_1 \mod L,
    \label{sumofzeros}
\end{equation}
where mod $L$ indicates that the equality should be taken modulo integer multiples of $L_1$ and $L_2$.
This constraint is derived by first inserting the boundary condition of the sigma function in Eq.~\eqref{eq:quasiperiodicitytorus} into the wave function in Eq.~\eqref{defsingleparticlewf} and then comparing to the boundary condition on $f(z)$ in Eq.~\eqref{defbclll}, which yields $\theta_i = -i \left( L_i^* \Box - L_i \Box^* \right)/N_\phi\mod 2\pi $. 
One then arrives at Eq.~\eqref{sumofzeros} by comparison to the equation for the torus area in Eq.~\eqref{eq:torusflux} written in complex coordinates, i.e., $L_1^*L_2 - L_1L_2^* = 2\pi i N_\phi \ell_B^2$.

\subsubsection{Bloch Landau level states}
\label{sec:LLlattice}

Our motivation for reviewing the LLL states on the torus is for comparison to Bloch bands and, in particular, ideal bands. Since Bloch's theorem prescribes that eigenstates in a Bloch band are classified by their translation properties, to complete the comparison requires the translation eigenstates in the LLL, which is a subtle question due to the magnetic field.

To apply Bloch's theorem requires a lattice of ``magnetic unit cells,'' each containing integer flux per cell, so that the translation operators commute.
While the quantum Hall problem has continuous translation symmetry, a magnetic unit cell can be introduced as follows:
for a torus with primitive vectors $\bm L_1$, $\bm L_2$ enclosing $N_\phi$ flux quanta, choose two integers $N_1N_2 = N_\phi$.
A lattice is formed by the $N_1 \times N_2$ grid with basis vectors $\bm a_1 = \bm L_1 /N_1$ and $\bm a_2 = \bm L_2/N_2$.
Since each unit cell encloses one flux quantum, $\bm a_1 \times \bm a_2 = 2\pi\ell_B^2$, and hence $t(\bm a_1)$ and $t(\bm a_2)$ commute, though their composition has a minus sign, 
\begin{equation}
    t(\bm a_1) t(\bm a_2) = t(\bm a_2) t(\bm a_1) = -t(\bm a_1 + \bm a_2), \label{magtranuc}
\end{equation}
a marked distinction from conventional translation symmetry.

We now define Bloch states for the LLL wave functions on the torus in Eq.~(\ref{defsingleparticlewf}).
Since the zeros uniquely determine the LLL wave function, and the magnetic translations $t(\bm a_i)$ shift the wave function -- and therefore the zero pattern -- uniformly by $\bm a_i$, the zeros of the Bloch states must be arranged on an equally spaced $N_1 \times N_2$ grid on the torus given by
\begin{equation}
    \omega_{\bm k}^{m,n} = a_o(\bm k) + m  a_1 + n  a_2; \quad m = 0,\dots N_1-1; \quad n = 0,\dots N_2-1,
\end{equation}
where $a_o(\bm k)$ is the origin of the grid of zeros for the Bloch state at $\bm{k}$.
The boundary condition in Eq.~\eqref{sumofzeros} fixes the sum of the zeros for every value of $\bm k$ to be:
\begin{equation}
	\sum_{m,n} \omega^{m,n}_{\bm k}  
	= N_\phi  a_o(\bm{k}) 
	= \frac{\theta_1}{2\pi } L_2 - \frac{\theta_2}{2\pi } L_1 \mod L,
\end{equation}
where we have assumed $N_1 = N_2 \mod 2$ (there is a small modification otherwise).
There are $N_\phi$ solutions to this equation, given by:
\begin{equation}
	a_o(\bm k) = \frac{\theta_1}{2\pi} \frac{a_2}{N_1} - \frac{\theta_2}{2\pi} \frac{a_1}{N_2} + z_k, \quad z_k = -ik \ell_B^2 
    \label{eq:origins}
\end{equation}
where the complexified Bloch momentum is given by 
\begin{equation}
    k \in \lbrace \frac{i}{\ell_B^2}\left( m \frac{a_2}{N_1} + n \frac{a_1}{N_2} \right) | m=0, \dots N_1-1, n=0, \dots N_2-1 \rbrace.
    \label{eq:kvalues}
\end{equation}
In the complex notation, the reciprocal lattice vectors are given by $b_{1,2} = -i a_{2,1}/\ell_B^2$ so that $\bm a_i \cdot \bm b_j = 2\pi \delta_{ij}$; thus Eq.~\eqref{eq:kvalues} is consistent with the usual notion of momentum on a finite-size torus.
The $N_\phi$ values of $k$ give rise to $N_\phi$ wave functions (defined momentarily), which span the $N_\phi$ linearly independent states in the LLL. 

The wave functions for the magnetic Bloch states can also be written in terms of a second set of sigma functions, which are quasi-periodic on the torus spanned by $\bm a_{1,2}$, satisfying
\begin{equation}
    \sigma(z+a_i) = - e^{a_i^*(z+\frac{a_i}{2})/\ell_B^2} \sigma(z).
\end{equation}
This boundary condition is identical to Eq.~\eqref{eq:quasiperiodicitytorus} with $(\bm L_1, \bm L_2, N_\phi) $ replaced by $(\bm a_1, \bm a_2, 1)$. The magnetic Bloch state wavefunction is then~\cite{JW_CTBG_Nodal},
\begin{equation}
    \Phi_{\bm k}(\bm r) = \sigma(z - z_k) e^{z_k^* z/\ell_B^2} e^{-\frac{1}{2}|z|^2/l_B^2} e^{-\frac{1}{2}|z_k|^2/l_B^2}, \label{lllwfform}
\end{equation}
where $z_k = -ik\ell_B^2$. The Gaussian normalization factor $\exp(-\bm k^2l_B^2/2)$ is inserted so that the wavefunction appears symmetric under coordinate and momentum exchange, which will be useful in the next section.
From the action of the translation operator in Eq.~\eqref{eq:translationLLL}, the magnetic Bloch state $\Phi_{\bm k}$ transforms as,
\begin{equation}
    \hat t(\bm a) \Phi_{\bm k} = \eta_{\bm a} e^{i\bm k\cdot\bm a} \Phi_{\bm k},
    \label{eq:LLtranslationphase}
\end{equation}
where the factor $\eta_{\bm a} = (-1)^{m+n+mn}$ for $\bm a = m \bm a _1 + n \bm a _2$ 
originates from Eq.~\eqref{magtranuc}. 
Aside from this phase factor, Eq.~\eqref{eq:LLtranslationphase} resembles Bloch's theorem.

We now return to the inhomogeneous magnetic field, $B(\bm r) = \bar B + \delta B(\bm r)$. Following the discussion in Sec.~\ref{sec:inhomogeneous}, the only effect of a real space modulation of the magnetic field to the wavefunction is a factor $e^{-e\phi(\bm r)/\hbar}$, where the Laplacian of $\phi(\bm r)$ yields the modulation $\delta B(\bm r)$ via Eq.~\eqref{eq:laplacianphi}. This establishes a bijective mapping between the Hilbert spaces corresponding to the homogeneous and inhomogeneous fields.
If the modulation is periodic on an $N_1 \times N_2$ lattice, the magnetic Bloch momentum $\bm k$ remains well defined, so that the magnetic Bloch state, which obeys the magnetic translation property in Eq.~\eqref{eq:LLtranslationphase}, can be written as
\begin{equation}
    \psi_{\bm k}(\bm r) = e^{-e\phi(\bm r)/\hbar} e^{-N(\bm k)} \Phi_{\bm k}(\bm r),\label{defpsik}
\end{equation}
where the term $e^{-N(\bm k)}$ ensures the wave function is normalized. 
From the resemblance of Eq.~\eqref{eq:LLtranslationphase} to the usual form of Bloch's theorem, 
Eq.~\eqref{defpsik} can be regarded as the ``Bloch state'' of the magnetic translation operator.

\subsubsection{Geometry of Bloch states}
We now introduce the ``cell periodic part'' of the magnetic Bloch function $u_{\bm k}(\bm r) \equiv e^{-i\bm k\cdot\bm r}\psi_{\bm k}(\bm r)$.
Using the definition of $z_k$ in Eq.~\eqref{eq:origins}, the Bloch factor $\exp(-i\bm k\cdot\bm r) = \exp[-(z_k^*z-z_kz^*)/l_B^2]$.
Combined with the wave function in Eq.~\eqref{defpsik}, 
\begin{equation}
    u_{\bm k}(\bm r) = -e^{-e\phi(\bm r)/\hbar} e^{-N(\bm k)} \Phi_{\bm r}(\bm k).
    \label{defuk}
\end{equation}
Importantly, comparing Eqs.~\eqref{defpsik} and \eqref{defuk}, the $\bm r$ and $\bm k$ indices on $\Phi_{\bm r}(\bm k)$ are exchanged. This reflects a position-momentum duality of the LLL wavefunction: the ``cell periodic function'' $u_{\bm k}(\bm r) \sim \Phi_{\bm r}(\bm k)$ and the ``Bloch function'' $\psi_{\bm k}(\bm r) \sim \Phi_{\bm k}(\bm r)$ are related by swapping position and momentum.

Aside from normalization factors, $u_{\bm k}(\bm r)$ depends only on the holomorphic part of the momentum, $z_k$.
As we will see in Sec.~\ref{sec:ideal}, this condition is equivalent to saturating the trace bound in Eq.~\eqref{eq:inequalities}, i.e., ${\rm Tr}g_{\bm k} = \Omega_{\bm k}$, and is the motivation for the definition of the ideal band.

We now derive the Berry curvature of the Bloch band. To utilize the holomorphic structure of the wave function in momentum space, we introduce the holomorphic $A_{\bm k} = w_a^* A^a_{\bm k}$ and anti-holomorphic $\bar A_{\bm k} = w_a A^a_{\bm k}$ Berry connections. The anti-holomorphic component is easier to compute, as it only depends on the normalization factor,
\begin{equation}
    \bar A_{k} \equiv w_a A^a_{\bm k} = -i\langle u_{\bm k} | \partial_{\bar k} u_{\bm k}\rangle = i \partial_{\bar k} N(\bm k) + i k \ell_B^2/2,
    \label{eq:BerryconnectionLLL}
\end{equation}
where we have introduced the holomorphic and anti-holomorphic momentum space derivatives, $\partial_k = w_a^* \partial_k^a$ and $\partial_{\bar k} = w_a \partial_k^a$. The holomorphic component of the connection is the complex conjugate of $\bar A$. This gives the Berry curvature~\cite{wang2021exact}:
\begin{equation}
    \Omega_{\bm k} = -i \partial_k \bar A_{\bm k} + i \partial_{\bar k} A_{\bm k} = \ell_B^2 + (\partial_k \partial_{\bar k} + \partial_{\bar k} \partial_{k}) N(\bm k) = \ell_B^2 + \Delta_{\bm k} N(\bm k),
    \label{eq:BerrycurvatureLLL}
\end{equation}
where $\Delta_{\bm k} = \partial_k \partial_{\bar k} + \partial_{\bar k} \partial_{k} = g_{ab} \partial_k^a \partial_k^b$ is the Laplacian in momentum space.

That the function $u_{\bm k}(\bm r)$ is written in the form of a LLL wavefunction on the torus but with the role of position and momentum swapped suggests that ideal bands establish a duality between real and momentum space. This is further supported by the computation of the Berry curvature in Eq.~\eqref{eq:BerrycurvatureLLL}, which shows that $\phi(\bm r)$ and $N(\bm k)$ should be viewed as controlling the modulation of the K\"ahler potentials in real and momentum space, respectively.

\subsection{Many-body wavefunctions}
\label{sec:manybodyLLL}

We now turn to many-body wave functions built from electrons in the LLL, taking the Laughlin state at filling $\nu = 1/m$ as an example. 
A crucial feature of a topologically ordered state is its ground state degeneracy, which depends on the genus of the compact manifold on which the problem is studied.

On the infinite plane -- which is topologically equivalent to a sphere and thus has a unique ground state -- the famous Laughlin wavefunction is written as \cite{laughlin1983anomalous},
\begin{equation}
    \Psi_{\rm Laughlin}(\{z_i\}) = \prod_{i<j} \left(z_i - z_j\right)^m \prod_{i} e^{-\frac12|z_i|^2/l_B^2},
    \label{eq:laughlin}
\end{equation}
where the parameters $z_{i=1,\dots N_e}$ specify the positions of $N_e$ particles.
Comparison to Eq.~\eqref{eq:psiLLL} shows that this many-body wave function is made up entirely of single-particle wave functions in the LLL.
The wavefunction also has a special clustering property: when two particles come close, the wavefunction vanishes as their separation to the $m^\text{th}$ power. This goes beyond the constraints of quantum statistics, which only require linear vanishing for fermions (and none for bosons). The enhanced clustering reflects the structure of the two-particle reduced density matrix and implies that the Laughlin state is the exact ground state of repulsive interactions that penalize pair states with relative angular momentum less than $m$. This ``parent Hamiltonian'' can be written as $\sum_{i<j} V(\bm r_i - \bm r_j)$, where \cite{trugman1985exact}
\begin{equation}
    V(\bm r) = \sum_{p=0,\dots,m-1} v_p ~ \mathcal P_{\rm LLL} \left[\nabla^{2p} \delta(\bm r)\right] \mathcal P_{\rm LLL}, \label{eq:interactionHam}
\end{equation}
$\mathcal P_{\rm LLL}$ denotes the projector into the LLL,
and the coefficients $v_p \geq 0$ are related to the Haldane pseudopotentials~\cite{haldane1983fractional}.
Only the odd (even) values of $p$ matter for fermions (bosons).
Due to the holomorphicity of the LLL, the index $p$ both counts the order of a zero when two particles coincide (clustering) and gives the relative angular momentum of the pair.

The Laughlin wavefunction on the plane generalizes to the torus as: 
\begin{equation}
    \Psi^{(m)}_{\rm Laughlin}(\{z_i\}; \bm\alpha) = \prod_{i<j} \tilde\sigma^m(z_i-z_j) \prod_{l=1}^{m} \tilde\sigma(\sum_i z_i - \alpha_l) \prod_{i} e^{-\frac12|z_i|^2/l_B^2}. \label{torusLaughlin}
\end{equation}
There are two modifications compared to the wave function on the plane: first, the holomorphic Jastrow factors are replaced by sigma functions, and second, a center of mass term is added.
In addition to accounting for the $m$-fold ground state degeneracy, the latter term ensures that each single particle state in the many-body wave function belongs to the LLL with the correct boundary conditions on the torus, as we now explain.

Following Sec.~\ref{sec:LLLwftorus}, to demonstrate that each single particle state belongs to the LLL, it must satisfy two conditions: (1) it has $N_\phi$ total zeros (counting an order-$m$ zero as $m$ zeros); and (2) the $N_\phi$ zeros satisfy the boundary condition in Eq.~\eqref{sumofzeros}. 
Now without loss of generality, consider the particle at $z_i$.
The wave function has $(N_e-1)$ zeros, each of order $m$, from the Jastrow factors $\tilde{\sigma}^m(z_i-z_j)$ and another $m$ order-one zeros from the center of mass term.
Recalling that the Laughlin wave function describes a state at filling $N_e/N_\phi = 1/m$, this does indeed add to $N_\phi$ total zeros; thus, the first condition is satisfied.
To address the second condition requires the sum of the positions of these zeros, $m\sum_{j\neq i}z_j + \sum_{l}\left( \alpha_l - \sum_{j\neq i} z_j \right) = \sum_l \alpha_l$.
There are $m$ linearly independent values of $\bm \alpha$ that satisfy the boundary condition in Eq.~\eqref{sumofzeros}.
Therefore, the wavefunction is a valid generalization to the torus and there are $m$ linearly independent ground states. Importantly, since the $\tilde{\sigma}(z)$ function is asymptotically equivalent to $z$ for small $|z|$, the Laughlin wavefunction on the torus enjoys identical clustering properties as on the plane.

In an inhomogeneous magnetic field, the generalized Laughlin wave function is identical to Eq.~\eqref{eq:laughlin} or \eqref{torusLaughlin} on the appropriate manifold, with an additional factor $\prod_{i} e^{-e\phi(\bm r_i)/\hbar}$. It remains a zero-energy ground state when the interaction term takes the form of Eq.~\eqref{eq:interactionHam}, because these additional factors do not modify the clustering property of the FQH states.

\section{IDEAL BANDS} 
\label{sec:ideal}
Searching for fractionalized phases in models or materials is intrinsically a difficult problem in many-body physics. One approach is to engineer a single-particle band to mimic a Landau level, hoping that fractionalized phases emerge following the paradigm of fractional quantum Hall (FQH) physics. To pursue this approach requires defining what it means for a Bloch band to resemble a Landau level. This section aims to review  efforts~\cite{parameswaran2012fractional,parameswaran2013fractional,roy2014band,ledwith2020fractional,wang2021exact,mera2021,mera2021b,ledwith2023vortexability,estienne2023ideal} towards providing ``Landau level mimicry criteria'' for Bloch bands.

The first challenge is that not all Landau levels are the same: while they share identical topology and Berry curvature, higher Landau levels have a larger quantum metric, i.e., larger $\operatorname{Tr} g$, reflecting that their wavefunctions are more spread out in space. This explains why charge-ordered states are favored in higher Landau levels. The LLL is special as it has the smallest quantum metric and saturates the quantum geometric bound in Eq.~\eqref{eq:inequalities}. 
This feature is reflected in its wave function: as discussed in Sec.~\ref{sec:LLLwftorus}, the Bloch LLL wave functions are holomorphic in the real-space coordinate $z$, while their periodic parts $e^{-i\bm k\cdot\bm r}\psi_{\bm k}(\bm r)$ exhibit momentum-space holomorphicity, depending only on $k$. 
We will show in this section that these conditions are equivalent: saturating the trace bound is equivalent to the momentum-space holomorphicity of cell-periodic wavefunctions.
These special features of the LLL motivate our focus on mimicking it.

Imitating the LLL is not new: 
Refs.~\cite{parameswaran2012fractional, parameswaran2013fractional,roy2014band} proposed mimicking the famous Girvin-MacDonald-Platzmann (GMP) algebra \cite{girvin1986magneto} of density operators in the LLL.
There it was shown that the GMP algebra of an isolated Bloch band is satisfied to third order in the momentum of the density operator when the Berry curvature is uniform and the trace bound in Eq.~\eqref{eq:inequalities} is saturated. 
Recently it was further proved that the Landau level is the only place where the algebra is exactly satisfied \cite{wang2025closed}. 

However, even satisfying the GMP algebra to third order may be too strict: for example, the Aharonov-Casher \cite{aharonov1979ground} model describing a $g=2$ particle in a spatially varying magnetic field (see Sec.~\ref{sec:inhomogeneous}) exhibits an exact zero-energy ground state that generalizes the Laughlin wave function for a short-range interaction. That ground state persists for arbitrarily large variations in magnetic field 
as long as the interaction remains ideal -- for a projected Coulomb interaction, Berry curvature fluctuations will eventually destroy the Laughlin ground state~\cite{shi2025effects}.
While this model can be far from the limit of uniform Berry curvature, it does saturate the trace bound in Eq.~\eqref{eq:inequalities}.
This further motivates the definition of an \emph{ideal band} as one that saturates the trace bound without imposing any constraint on Berry curvature uniformity.  
In the following, we review properties of such ideal bands and their wave functions.

\subsection{Trace condition and holomorphicity}\label{sec:tracecondholo}
We define an ideal band as one where the trace bound in Eq.~\eqref{eq:inequalities}, or equivalently, Eq.~\eqref{eq:Chernbound}, is saturated for a $\bm{k}$-independent choice of metric, i.e., there exists a $\bm{k}$-independent metric $h_{ab}$, which defines a complex structure via Eq.~\eqref{defcomplexstructure}, such that
\begin{equation}
     \text{Tr}_h g_{\bm k} \equiv h_{ab} g^{ab}_{\bm k}  = |\Omega_{\bm k}|,\quad \forall \mathbf \, \bm k.\label{deftracecond}
\end{equation}
We refer to Eq.~\eqref{deftracecond} as saturating the \emph{trace condition}. 

A \emph{Kähler band} is defined as one whose cell-periodic part of the Bloch wave function depends on momentum only through its holomorphic component $k = w^a(\bm k) k_a$, up to a normalization factor; it saturates the second inequality (the so-called \emph{determinant bound}) in Eq.~\eqref{eq:inequalities}. 

The following conditions are equivalent:
\begin{enumerate}
     \item The band $|u_{\bm{k}} \rangle$ is an ideal band, saturating the trace condition.
     \item The QGT of $|u_{\bm{k}} \rangle$ has a $\bm{k}$-independent null vector.
     \item The band $|u_{\bm{k}}\rangle$ is a Kähler band with $\bm{k}$-independent complex structure.
\end{enumerate}
More concisely, an ideal band can be written as,
\begin{equation}
    |u_{\bm k}\rangle = \frac{1}{\mathcal N_{\bm k}} |\tilde u_{k}\rangle,\label{defholomorphicity}
\end{equation}
where $|u_{\bm k}\rangle$ denotes the normalized cell-periodic part of the Bloch wave function, $\mathcal{N}_{\bm k}$ is a real normalization factor, and $|\tilde u_{k}\rangle$ depends on the momentum only through the holomorphic component of the momentum coordinate, so that $\partial_{\bar k}|\tilde u_{k}\rangle = 0$.

The equivalence of conditions 1. and 2. is reviewed in the SM Sec.~\ref{sec:QGTproofs}.
That holomorphicity (condition 3) implies the trace bound (conditions 1, 2) was noticed in early literature such as Ref.~\cite{claassen2015position}. We provide a quick proof following Ref.~\cite{ledwith2020fractional}. The starting point is to rewrite the QGT in terms of the unnormalized state $|\tilde{u}_k\rangle$ as 
\begin{equation}
    Q^{ab}_{\bm k} = \frac{\langle\partial_k^a\tilde u_k|\partial_k^b\tilde u_k\rangle}{\mathcal N_{\bm k}^2} - \frac{\langle\partial_k^a\tilde u_k|\tilde u_k\rangle\langle\tilde u_k|\partial_k^b\tilde u_k\rangle}{\mathcal N_{\bm k}^4}.
    \label{eq:QGTunnormalized}
\end{equation}
By virtue of being a holomorphic band, $\partial_{\bar k}|\tilde u_k\rangle = 0$ (recall $\partial_{\bar{k}} = w_b\partial^b_k$). Thus, Eq.~\eqref{eq:QGTunnormalized} implies $Q^{ab}_{\bm k}w_b=0$, so that the QGT has a constant null vector.

In the other direction, that saturating the quantum geometrical bound implies the holomorphicity of Bloch states was proved in Ref.~\cite{mera2021,mera2021b} by showing that a Bloch state, formally regarded as a mapping from the BZ torus to complex projective space, is a holomorphic immersion if and only if the determinant bound (second inequality in Eq.~\eqref{eq:inequalities}) is saturated, motivating the definition of the K\"ahler band above.
An ideal band is a special class of K\"ahler band whose complex structure is constant across the BZ. 
Quasi-K\"ahler bands are K\"ahler bands where the Berry curvature vanishes at at least one momentum, which is the case for two-band models in 2D~\cite{mera2021}.

\emph{Vortexability} provides a real-space perspective on ideal bands and beyond~\cite{ledwith2023vortexability}. Defining $\mathcal P$ as the band projector and $1 - \mathcal P$ as its complement, the vortexability criterion states that there exists a vortex operator $z(\mathbf r)$ associated with every point in real space such that $(1-\mathcal P) z(\mathbf r) \mathcal P = 0$. In an ideal band, the vortex operator is simply the holomorphic component of the position operator $z = w_a r^a$, and vortexability illustrates the real-space holomorphic properties of ideal bands. Such a real-space perspective
enables the direct construction of exact FCI ground states for
ideal bands with short-range interactions~\cite{ledwith2023vortexability,Dong_decomposition23}; see Sec.~\ref{sec:FCI}. Vortexability also implies perfect circular magnetic dichroism in optical responses~\cite{zerofieldCFL23}.

\begin{figure}
    \includegraphics[width = 0.5\textwidth]{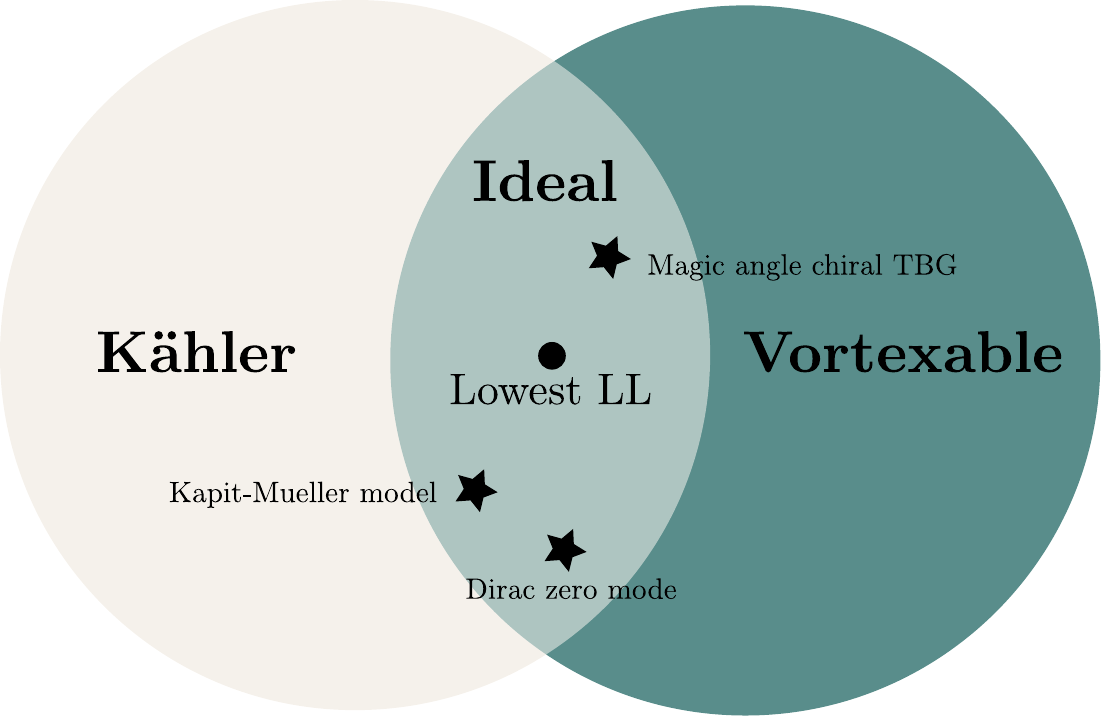}
    \caption{The relationship between K\"ahler, ideal, and vortexable bands. K\"ahler and vortexable bands are defined by momentum and real space holomorphicity, respectively, while ideal bands lie at their intersection. Figure adopted from Ref.~\cite{liu2025theory}.}\label{fig_bands}
\end{figure}

A comparison of K\"ahler, ideal and vortexable bands is summarized in Fig.~\ref{fig_bands}.
The LLL is the classic example of an ideal band. 
More examples and their experimental relevance are discussed in Sec.~\ref{sec:examples}. In the SM Sec.~\ref{sec:kapitmueller} we provide an example of a vortexable but non-ideal band.

\subsection{From holomorphicity to ideal band wavefunctions}

We now seek an explicit form of the wave function of an ideal band. 
As discussed in Sec.~\ref{sec:geometry}, a non-zero Chern number implies the absence of a globally defined smooth gauge choice for Bloch wave functions. In the following we assume a gauge that is smooth inside of the BZ torus, which implies a discontinuity at the BZ boundary. 

As discussed in the previous section, the trace condition \emph{necessarily} and \emph{sufficiently} ensures the holomorphicity of the Bloch state, i.e., the state can be written in the form of Eq.~\eqref{defholomorphicity}.
As we learned from the LLL wavefunction that real space holomorphicity strongly constrains the real space wavefunction (see Sec.~\ref{sec:LLLwftorus}), similarly, momentum space holomorphicity strongly constrains the analytical form of ideal band Bloch wavefunctions.

\begin{figure}
    \includegraphics[width = \textwidth]{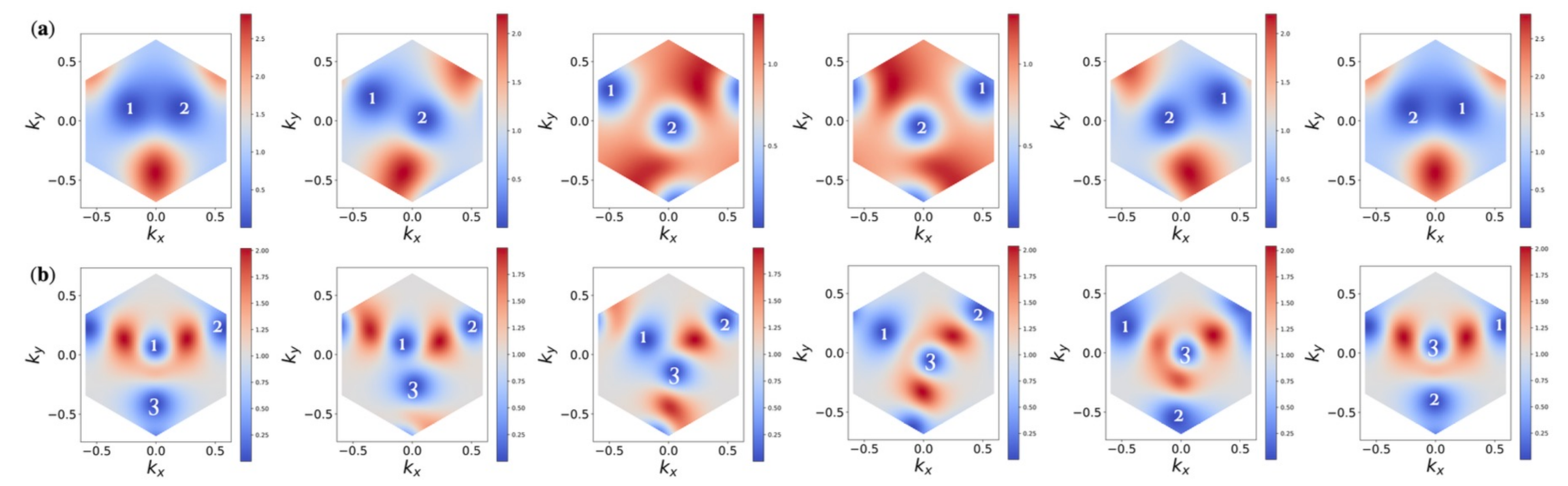}
    \caption{Cell-periodic part of the Bloch wave function of an ideal band with Chern number $C=2$ in panel (a) and Chern number $C=3$ in panel (b). Each column is a plot of $|u_{\bm k}(\mathbf r)|$ as a function of $\bm k$ for different choices of $\mathbf r$, sampled linearly on a line from the initial point $\mathbf r_0$ to the final point $\mathbf r_0 + \mathbf a$, where $\mathbf a$ is a lattice vector. The center of the $\bm k-$space zeros are evolving linearly in the BZ following the change of $\mathbf r$. Figure adopted from Ref.~\cite{wang2023origin}.}\label{fig_idealband_zeros}
\end{figure}

To understand the constraint, we start by comparing the periodic functions at momenta related by reciprocal lattice vectors. The states must obey,
\begin{equation}
    \tilde{u}_{k+b_i}(\bm r) = e^{i\theta_i(\bm r)} e^{i\phi_i(k)} \tilde{u}_k(\bm r),\quad \theta_i \equiv -\bm b^i\cdot\bm r, \label{kspacebc}
\end{equation}
where we have used $b^i$ to denote the complex reciprocal lattice vector.
The condition parallels the boundary condition on the holomorphic part of the LLL wave function on the torus in Eq.~\eqref{defbclll}:
the phase $\phi_i(k)$ is required to be holomorphic and the phase $\exp(-i\bm b_i\cdot\bm r)$ is dual to the fictitious flux appearing in Eq.~\eqref{defbclll}. 
Mathematically, the analog to the contour integral in Eq.~\eqref{eq:Nphizeros} arises because for any real space point $\bm r$, the Chern number reflects the boundary integral of $\tilde u_k$ under a smooth choice of gauge, yielding
\begin{equation}
    C = \frac{1}{2\pi i} \oint dk ~ \partial_k \ln \tilde{u}_k(\bm r). \label{kspacecauchy}
\end{equation}
This suggests that in an ideal band $\tilde u_k(\bm r)$, one should view coordinate $\bm r$ as tuning the `fictitious flux' determining the dual boundary condition, parallel to  $(\theta_1, \theta_2)$ in the quantum Hall problem, and one should regard the Chern number $C$ as the `total flux quanta', parallel to $N_\phi$ in the quantum Hall problem~\cite{wang2023origin}. 
This comparison can be made exact.
The model calculations in Fig.~\ref{fig_idealband_zeros} show the wave function of an ideal band, which indeed has $C$ zeros when plotted in $\bm k-$space for fixed $\bm r$. The $C$ zeros must sum to the real space coordinate $\bm r$, paralleling the constraint on the sum of zeros in Eq.~\eqref{sumofzeros}.

The ideal band wave function can be obtained in the same spirit as the quantum Hall wave function on the torus: holomorphicity ensures that once a wave function is found, it is unique. In the $C=1$ case, the ideal band wavefunction must take the form \cite{wang2021exact},
\begin{equation}
    \psi_{\bm k}(\mathbf r) = e^{-N(\bm k)} \mathcal B(\mathbf r) \Phi_{\bm k}(\mathbf r), \label{defidealbandwf1}
\end{equation}
where $N(\bm k)$ is the normalization factor that accounts for the variation of Berry curvature in momentum space, and $\mathcal B(\mathbf r)$ is a momentum independent factor. The wave function in Eq.~\eqref{defidealbandwf1} is quite similar to the Aharonov-Casher wavefunction discussed in Eq.~\eqref{defpsik}, except with the exponential of the scalar field $\exp[-e\phi(\mathbf r)/\hbar]$ in Eq.~\eqref{defpsik} replaced by $\mathcal B(\mathbf r)$. 
Notice that the wave function in Eq.~\eqref{defidealbandwf1} is still a Bloch wave function, which  obeys the standard translation properties of Bloch bands, in contrast to the magnetic translation properties of states in the LLL. Thus, the factor $\mathcal B(\mathbf r)$ must be quasi-periodic in the real space to compensate for the quasi-periodicity of the LLL wavefunction $\Phi_{\bm k}(\mathbf r)$.

For a more general $C>1$ ideal band, the wavefunction is given by ~\cite{wang2023origin,Dong_decomposition23},
\begin{equation}
    \psi_{\bm k}(\bm r) = e^{-N(\bm k)} \sum_{\sigma = 1}^{C} \mathcal{B}(\bm r+\sigma\bm a_1) \Phi_{\bm k}(\bm r+\sigma\bm a_1). \label{defidealbandwf2}
\end{equation}

The ideal band has also been generalized to higher Landau levels and applied to fractionalization in moir\'e systems~\cite{liu2025theory,HigherVortexability25,FCWu_VariatoinalMapping25,FCWu_GLL26}.

\subsection{Implication of ideal bands for fractional Chern insulators}
\label{sec:FCI}

An important property of ideal bands is that model FCI wave functions are exact zero energy ground states at the appropriate filling and with short ranged interactions. 
Here we focus on $C=1$, but the statement extends to ideal bands with $C>1$~\cite{wang2023origin,Dong_decomposition23}.

For a $C=1$ ideal band, the ideal quantum geometry implies an isomorphism to the LLL states, as shown by Eqs.~\eqref{defidealbandwf1} and \eqref{defpsik}.
Hence, the many-body wave functions are also in one-to-one correspondence.
Thus, a many-body quantum Hall wave function in the LLL, immediately yields a corresponding generalized FCI wave function in a particular ideal band.
For instance, the generalized $\nu = 1/m$ Laughlin state in an ideal band is,
\begin{equation}
    \Psi^{(m)}_{\rm ideal}(\{z_i\}, \bm\alpha) = \left[\prod_i \mathcal B(\bm r_i)\right] \Psi^{(m)}_{\rm Laughlin}(\{z_i\}, \bm\alpha),
\end{equation}
where $\Psi^{(m)}_{\rm Laughlin}(\{z_i\}, \bm\alpha)$ is the LLL Laughlin wave function in Eq.~\eqref{torusLaughlin}. From the wave function in Eq.~\eqref{defidealbandwf1}, $\Psi^{(m)}_{\rm ideal}$ is within the Fock space determined by the ideal band. Additionally, from the form of the Laughlin wave function, as $z = z_i - z_j \rightarrow 0$, the wave function vanishes as $z^m$ and thus the probability density vanishes as $|z|^{2m}$. Hence, following Ref.~\cite{ledwith2020fractional,ledwith2023vortexability}, the interaction energy of $\Psi^{(m)}_{\rm Laughlin}$ vanishes for any two-body interaction term of the form of Eq.~\eqref{eq:interactionHam}.

\subsection{Examples of ideal bands and experimental relevance} \label{sec:examples}

So far, we have provided two examples of ideal bands: the LLL ($n=0$) and the Aharonov-Casher band. 
While individual higher $n\geq1$ Landau levels are not ideal, the many-body state formed by filling the lowest $p \geq 0$ Landau levels together yield an ideal band. This essential fact enables the generalization of ideal bands to higher Landau levels~\cite{liu2025theory,HigherVortexability25}.

Beyond Landau levels and Aharonov-Casher model, ideal bands are now realized in many lattice systems.
In particular, ideal bands have attracted significant attention due to their relevance to twisted bilayer graphene: in the ``chiral model'' of twisted bilayer graphene \cite{tarnopolsky2019origin} at the magic twist angle, the system exhibits two exactly flat bands, each localized on a different sublattice and each realizing an ideal band with Chern number $C=1$. As a result, exact fractional Chern insulator states are theoretically predicted in this limit~\cite{ledwith2020fractional}. Although in realistic twisted bilayer graphene systems, stabilizing fractionalized phases typically requires a finite magnetic field \cite{xie2021fractional}, the exact results in the chiral limit provide important insights into the emergence of fractionalized phases in twisted materials.

The first zero field FCI was observed in moir\'e MoTe$_2$ \cite{cai2023signatures,zeng2023thermodynamic,park2023observation, xu2023observation}. Theoretical calculations based on both first principle and continuum models indicate that the narrow band in such systems resembles a LLL in many aspects, including the nearly uniform Berry curvature distribution, near-ideal quantum geometry and narrow band dispersions \cite{wu2019topological,devakul2021magic,li2021spontaneous,crepel2023anomalous,morales2023pressure,jia2024moire,crepel2024chiral,wang2025higher}. 
Later theoretical work proposed that the origin of such LLL-like flat bands arises from real-space skyrmion winding: the spatial winding of the layer pseudospin generates an effective magnetic field experienced by electrons that follow this texture~\cite{morales2024magic}. This provides an intuitive explanation for the emergence of nearly ideal bands in twisted MoTe$_2$ within the Aharonov–Casher framework~\cite{Shi_ACband_tTMD24}.

The second zero field FCI was observed in rhombohedral graphene with aligned hexagonal boron nitride \cite{lu2024fractional,xie2025tunable}, for which ideal band models have also been developed \cite{bernevig2025berry,tan2025ideal}.

Beyond twisted materials, ideal bands or near-ideal bands are predicted in periodically strained \cite{gao2023untwisting,wan2023topological} and nano-patterned materials \cite{ghorashi2023topological,zeng2024gate,ault2026optimizing}.

Finally, ideal bands can be realized in tight-binding models. A notable example is the Kapit-Mueller model~\cite{KapitMuellerModel}, where the hopping parameters are carefully engineered to achieve a positive semi-definite spectrum with an exactly flat band. Remarkably, the wavefunctions of this flat band take the form of lattice analogs of LLL states. In this sense, the Kapit-Mueller model provides a tight-binding realization of an ideal band, distinguishing it from continuum constructions such as the Aharonov-Casher approach or moiré-based systems described above. We briefly review this model in the SM Sec.~\ref{sec:kapitmueller}.

\section{BEYOND IDEAL BANDS}

\subsection{Deviations from ideality}
\label{sec:deviations}

Given the importance of ideal bands for FCIs, it is desirable to develop criteria to measure the deviations from ideal bands. 
To this end, two geometric indicators have been most commonly studied: the first is standard deviation of the Berry curvature: 
\begin{equation}
    F = \left[ \int_{\text{BZ}} \frac{d^2\bm{k}}{A_{\text{BZ}}} \left( C- \Omega_{\bm{k}}\frac{A_{\text{BZ}}}{2\pi} \right)^2 \right]^{1/2},
    \label{eq:defF}
\end{equation}
which was first introduced in the context of mimicking the GMP algebra \cite{parameswaran2012fractional} (see Sec.~\ref{sec:ideal}).
The second indicator is the deviation from the ideal band condition:
\begin{equation}
    \bar{T} = {\text{min}}_\omega \int_\text{BZ} \frac{d^2\bm{k}}{A_\text{BZ}}\left( \text{Tr}_\omega g_{\bm{k}} - |\Omega_{\bm{k}}| \right),
    \label{eq:defT}
\end{equation}
i.e., the deviation from the ideal band condition minimized over all choices of the matrix $\omega$, which was introduced in Ref.~\cite{morales2023pressure}, motivated by Refs.~\cite{roy2014band,jackson2015geometric}.
The bounds in Eq.~\eqref{eq:inequalities} imply that $\bar{T}\geq 0$.

Minimizing $F$ and $\bar{T}$ has been shown to be correlated with the stability of an FCI ground state~\cite{jackson2015geometric,bauer2016quantum,parker2021field,morales2023pressure,wu2024quantum}.
However, these single-particle properties are not sufficient to determine the existence of an FCI: the many-body ground state fundamentally depends on the details of the interactions. The motivation for considering such indicators is that an FCI in an ideal band represents a special point within a broader topological phase that can persist under small deformations of the Hamiltonian. Yet, even in an ideal band, the model FCI wave function discussed in Sec.~\ref{sec:FCI} is an exact ground state only for the short-range interactions specified in Eq.~\eqref{eq:interactionHam}. Indeed, modifying the interaction range~\cite{emanuel2025unifying} or introducing Berry curvature fluctuations~\cite{shi2025effects}, even while keeping the band geometry fixed, can destabilize the FCI ground state.

\subsection{Orbital embedding}
\label{sec:embeddingmodel}

A further limitation of the deviations introduced in Sec.~\ref{sec:deviations} arises because the QGT is not invariant under the choice of orbital embedding, as mentioned in Sec.~\ref{sec:embedding}.
(See Ref.~\cite{simon2020contrasting} for a detailed and pedagogical discussion.)
Thus, despite their success, the geometrical indicators described in Sec.~\ref{sec:deviations} cannot alone be reliable indicators of an FCI ground state, which again reflects the fundamental fact that a many-body ground state cannot be fully determined from single-particle physics alone.
In the SM Sec.~\ref{sec:kapitmueller}, we demonstrate the importance of embedding dependence in the Kapit–Mueller model by numerically studying the effect of embedding variations on the exact Laughlin $\nu = 1/2$ state stabilized by on-site repulsion. We find that changing the embedding destroys the single-particle ideal band condition, while leaving both the single-particle band structure and the many-body spectrum unchanged. Ultimately, the embedding change converts an ideal flat band into a vortexable non-ideal flat band, underscoring the sensitivity of band geometry to the embedding despite the invariance of the spectrum.

\subsection{Fractional Chern insulators far from the lowest Landau level}
We briefly mention the important observation that FCIs can exist in bands far from the LLL or ideal band limit \cite{bauer2022fractional,andrews2024stability}, including in bands with vanishing Chern number \cite{simon2015fractional,kourtis2018symmetry,lin2025fractional,liu2025topological}. 
Whether an analog of ideal bands exists for these cases remains an open question.

\section{OUTLOOK}
The developments reviewed here point toward a unifying perspective in which quantum geometry plays a central role in organizing and stabilizing fractional topological phases in Bloch bands. In particular, the emergence of momentum-space holomorphicity and the saturation of geometric bounds suggest that a class of Bloch bands can faithfully emulate the Hilbert space structure of Landau levels, providing a promising route toward realizing fractionalized phases in vanishing magnetic field.

Several open questions remain. A key challenge is to establish robust and broadly applicable criteria that relate geometric properties of Bloch bands to the stability of interacting phases. While the trace condition and related bounds offer valuable guidance, their precise role in realistic systems, where band dispersion, disorder, and multiband effects are unavoidable, remains to be fully understood. Extending these ideas beyond idealized limits, and clarifying the extent to which the holomorphic structure can persist or be approximated in experimentally relevant settings, is an important direction for future work.
On the experimental front, rapidly advancing platforms such as moiré materials provide unprecedented opportunities to engineer and probe quantum geometry in situ. The ability to tune band structure, interactions, and symmetry suggests that geometric optimization may become a practical design principle for stabilizing fractionalized phases. In this context, developing direct probes of the quantum metric and its interplay with Berry curvature will be crucial for connecting theoretical criteria to observable signatures.
More broadly, the interplay between geometry, topology, and interactions continues to reveal new structures beyond those rooted in Landau levels. Understanding how these elements combine to produce robust many-body phases -- potentially in settings without direct analogues in continuum quantum Hall systems -- remains an exciting frontier.

\section*{DISCLOSURE STATEMENT}
The authors are not aware of any affiliations, memberships, funding, or financial holdings that
might be perceived as affecting the objectivity of this review. 

\section*{ACKNOWLEDGMENTS}
The authors are grateful for fruitful conversations and collaborations with Dathan Ault-McCoy, Lei Chen, Valentin Cr\'epel, Junkai Dong, Nicolás Morales-Dur\'an, Liang Fu, Daniele Guerci, Duncan Haldane, Guangyue Ji, Eslam Khalaf, Semyon Klevtsov, Patrick Ledwith, Nabil Lhachemi, Zhao Liu, Bruno Mera, Andrew Millis, Tomoki Ozawa, David Palomino, Daniel Parker, Xin Shen, Kai Sun, Ashvin Vishwanath, Fengcheng Wu, Di Xiao, Bo Yang, Kang Yang and Wang Yao. The authors are especially grateful to Nabil Lhachemi, Zhao Liu and Tomoki Ozawa for their feedback on this manuscript and to Jinjie Zhang for producing Supplemental Fig. 1.
J.C. acknowledges support from the Flatiron Institute, a division of the Simons Foundation.

\bibliographystyle{ar-style4}
\bibliography{QGrefs}

@article{wang2021exact,
  title={Exact {Landau} level description of geometry and interaction in a flatband},
  author={Wang, Jie and Cano, Jennifer and Millis, Andrew J and Liu, Zhao and Yang, Bo},
  journal={Phys. Rev. Lett.},
  volume={127},
  number={24},
  pages={246403},
  year={2021},
  publisher={APS}
}

@article{morales2023pressure,
  title = {Pressure-enhanced fractional Chern insulators along a magic line in moir\'e transition metal dichalcogenides},
  author = {Morales-Dur\'an, Nicol\'as and Wang, Jie and Schleder, Gabriel R. and Angeli, Mattia and Zhu, Ziyan and Kaxiras, Efthimios and Repellin, C\'ecile and Cano, Jennifer},
  journal = {Phys. Rev. Res.},
  volume = {5},
  issue = {3},
  pages = {L032022},
  numpages = {6},
  year = {2023},
  month = {Aug},
  publisher = {American Physical Society},
  doi = {10.1103/PhysRevResearch.5.L032022},
  url = {https://link.aps.org/doi/10.1103/PhysRevResearch.5.L032022}
}

@article{claassen2015position,
  title={Position-momentum duality and fractional quantum {Hall} effect in {Chern} insulators},
  author={Claassen, Martin and Lee, Ching Hua and Thomale, Ronny and Qi, Xiao-Liang and Devereaux, Thomas P},
  journal={Phys. Rev. Lett.},
  volume={114},
  number={23},
  pages={236802},
  year={2015},
  publisher={APS}
}

@article{liu2025theory,
  title = {Theory of Generalized Landau Levels and Its Implications for Non-Abelian States},
  author = {Liu, Zhao and Mera, Bruno and Fujimoto, Manato and Ozawa, Tomoki and Wang, Jie},
  journal = {Phys. Rev. X},
  volume = {15},
  issue = {3},
  pages = {031019},
  numpages = {39},
  year = {2025},
  month = {Jul},
  publisher = {American Physical Society},
  doi = {10.1103/1zg9-qbd6},
  url = {https://link.aps.org/doi/10.1103/1zg9-qbd6}
}

@article{parameswaran2012fractional,
  title={Fractional {Chern} insulators and the $W_\infty$ algebra},
  author={Parameswaran, SA and Roy, R and Sondhi, Shivaji L},
  journal={Phys. Rev. B},
  volume={85},
  number={24},
  pages={241308},
  year={2012},
  publisher={APS}
}

@article{wang2023origin,
  title={Origin of model fractional {Chern} insulators in all topological ideal flatbands: Explicit color-entangled wave function and exact density algebra},
  author={Wang, Jie and Klevtsov, Semyon and Liu, Zhao},
  journal={Phys. Rev. Res.},
  volume={5},
  number={2},
  pages={023167},
  year={2023},
  publisher={APS}
}

@article{roy2014band,
  title={Band geometry of fractional topological insulators},
  author={Roy, Rahul},
  journal={Phys. Rev. B},
  volume={90},
  number={16},
  pages={165139},
  year={2014},
  publisher={APS}
}

@article{ledwith2020fractional,
  title={Fractional Chern insulator states in twisted bilayer graphene: An analytical approach},
  author={Ledwith, Patrick J and Tarnopolsky, Grigory and Khalaf, Eslam and Vishwanath, Ashvin},
  journal={Phys. Rev. Res.},
  volume={2},
  number={2},
  pages={023237},
  year={2020},
  publisher={APS}
}

@article{girvin1986magneto,
  title={Magneto-roton theory of collective excitations in the fractional quantum Hall effect},
  author={Girvin, SM and MacDonald, AH and Platzman, PM},
  journal={Phys. Rev. B},
  volume={33},
  number={4},
  pages={2481},
  year={1986},
  publisher={APS}
}

@article{jackson2015geometric,
  title={Geometric stability of topological lattice phases},
  author={Jackson, Thomas S and M{\"o}ller, Gunnar and Roy, Rahul},
  journal={Nat. Comm.},
  volume={6},
  number={1},
  pages={8629},
  year={2015},
  publisher={Nature Publishing Group UK London}
}

@article{simon2020contrasting,
  title={Contrasting lattice geometry dependent versus independent quantities: Ramifications for Berry curvature, energy gaps, and dynamics},
  author={Simon, Steven H and Rudner, Mark S},
  journal={Phys. Rev. B},
  volume={102},
  number={16},
  pages={165148},
  year={2020},
  publisher={APS}
}

@article{yu2025quantum,
  title={Quantum geometry in quantum materials},
  author={Yu, Jiabin and Bernevig, B Andrei and Queiroz, Raquel and Rossi, Enrico and T{\"o}rm{\"a}, P{\"a}ivi and Yang, Bohm-Jung},
  journal={npj Quantum Materials},
  volume={10},
  number={1},
  pages={101},
  year={2025},
  publisher={Nature Publishing Group UK London}
}

@article{parameswaran2013fractional,
  title={Fractional quantum Hall physics in topological flat bands},
  author={Parameswaran, Siddharth A and Roy, Rahul and Sondhi, Shivaji L},
  journal={Comptes Rendus Physique},
  volume={14},
  number={9-10},
  pages={816--839},
  year={2013},
  publisher={Elsevier}
}

@article{dobardvzic2014effective,
  title={Effective description of Chern insulators},
  author={Dobard{\v{z}}i{\'c}, Edib and Dimitrijevi{\'c}, M and Milovanovi{\'c}, MV},
  journal={Phys. Rev. B},
  volume={89},
  number={23},
  pages={235424},
  year={2014},
  publisher={APS}
}

@article{mera2021,
  title = {K\"ahler geometry and Chern insulators: Relations between topology and the quantum metric},
  author = {Mera, Bruno and Ozawa, Tomoki},
  journal = {Phys. Rev. B},
  volume = {104},
  issue = {4},
  pages = {045104},
  numpages = {13},
  year = {2021},
  month = {Jul},
  publisher = {American Physical Society},
  doi = {10.1103/PhysRevB.104.045104},
  url = {https://link.aps.org/doi/10.1103/PhysRevB.104.045104}
}

@ARTICLE{mera2021b,
       author = {{Ozawa}, Tomoki and {Mera}, Bruno},
        title = "{Relations between topology and the quantum metric for Chern insulators}",
      journal = {Phys. Rev. B},
         year = 2021,
        month = jul,
       volume = {104},
       number = {4},
          eid = {045103},
        pages = {045103},
          doi = {10.1103/PhysRevB.104.045103},
archivePrefix = {arXiv},
       eprint = {2103.11582},
 primaryClass = {cond-mat.mes-hall},
       adsurl = {https://ui.adsabs.harvard.edu/abs/2021PhRvB.104d5103O}
}

@article{tarnopolsky2019origin,
  title = {Origin of Magic Angles in Twisted Bilayer Graphene},
  author = {Tarnopolsky, Grigory and Kruchkov, Alex Jura and Vishwanath, Ashvin},
  journal = {Phys. Rev. Lett.},
  volume = {122},
  issue = {10},
  pages = {106405},
  numpages = {6},
  year = {2019},
  month = {Mar},
  publisher = {American Physical Society},
  doi = {10.1103/PhysRevLett.122.106405},
  url = {https://link.aps.org/doi/10.1103/PhysRevLett.122.106405}
}

@article{crepel2024chiral,
  title={Chiral limit and origin of topological flat bands in twisted transition metal dichalcogenide homobilayers},
  author={Cr{\'e}pel, Valentin and Regnault, Nicolas and Queiroz, Raquel},
  journal={Comm. Phys.},
  volume={7},
  number={1},
  pages={146},
  year={2024},
  publisher={Nature Publishing Group UK London}
}

@article{morales2024magic,
  title = {Magic Angles and Fractional {Chern} Insulators in Twisted Homobilayer Transition Metal Dichalcogenides},
  author = {Morales-Dur\'an, Nicol\'as and Wei, Nemin and Shi, Jingtian and MacDonald, Allan H.},
  journal = {Phys. Rev. Lett.},
  volume = {132},
  issue = {9},
  pages = {096602},
  numpages = {6},
  year = {2024},
  month = {Mar},
  publisher = {American Physical Society},
  doi = {10.1103/PhysRevLett.132.096602},
  url = {https://link.aps.org/doi/10.1103/PhysRevLett.132.096602}
}

@article{andrews2024stability,
  title = {Stability of fractional Chern insulators with a non-Landau level continuum limit},
  author = {Andrews, Bartholomew and Raja, Mathi and Mishra, Nimit and Zaletel, Michael P. and Roy, Rahul},
  journal = {Phys. Rev. B},
  volume = {109},
  issue = {24},
  pages = {245111},
  numpages = {14},
  year = {2024},
  month = {Jun},
  publisher = {American Physical Society},
  doi = {10.1103/PhysRevB.109.245111},
  url = {https://link.aps.org/doi/10.1103/PhysRevB.109.245111}
}

@article{bauer2022fractional,
  title = {Fractional Chern insulators with a non-Landau level continuum limit},
  author = {Bauer, David and Talkington, Spenser and Harper, Fenner and Andrews, Bartholomew and Roy, Rahul},
  journal = {Phys. Rev. B},
  volume = {105},
  issue = {4},
  pages = {045144},
  numpages = {17},
  year = {2022},
  month = {Jan},
  publisher = {American Physical Society},
  doi = {10.1103/PhysRevB.105.045144},
  url = {https://link.aps.org/doi/10.1103/PhysRevB.105.045144}
}

@article{lin2025fractional,
  title={Fractional {Chern} insulator states in an isolated flat band of zero {Chern} number},
  author={Lin, Zuzhang and Lu, Hongyu and Yang, Wenqi and Zhai, Dawei and Yao, Wang},
  journal={Preprint arXiv:2505.09009},
  year={2025}
}

@article{simon2015fractional,
  title = {Fractional {Chern} insulators in bands with zero Berry curvature},
  author = {Simon, Steven H. and Harper, Fenner and Read, N.},
  journal = {Phys. Rev. B},
  volume = {92},
  issue = {19},
  pages = {195104},
  numpages = {7},
  year = {2015},
  month = {Nov},
  publisher = {American Physical Society},
  doi = {10.1103/PhysRevB.92.195104},
  url = {https://link.aps.org/doi/10.1103/PhysRevB.92.195104}
}

@article{emanuel2025unifying,
  title = {Unifying framework for fractional Chern insulator stabilization},
  author = {Emanuel, Peleg and Keselman, Anna and Oreg, Yuval},
  journal = {Phys. Rev. B},
  volume = {112},
  issue = {23},
  pages = {235133},
  numpages = {9},
  year = {2025},
  month = {Dec},
  publisher = {American Physical Society},
  doi = {10.1103/9hpw-kz7g},
  url = {https://link.aps.org/doi/10.1103/9hpw-kz7g}
}

@article{zhang2025beyond,
  title={Beyond the Lowest {Landau} Level: Unlocking More Robust Fractional States Using Flat {Chern} Bands with Higher Vortexability},
  author={Zhang, Yitong and Sarkar, Siddhartha and Wan, Xiaohan and Parker, Daniel E and Lin, Shi-Zeng and Sun, Kai},
  journal={Preprint arXiv:2510.22831},
  year={2025}
}

@article{HigherVortexability25,
  title = {Higher Vortexability: Zero-Field Realization of Higher Landau Levels},
  author = {Fujimoto, Manato and Parker, Daniel E. and Dong, Junkai and Khalaf, Eslam and Vishwanath, Ashvin and Ledwith, Patrick},
  journal = {Phys. Rev. Lett.},
  volume = {134},
  issue = {10},
  pages = {106502},
  numpages = {8},
  year = {2025},
  month = {Mar},
  publisher = {American Physical Society},
  doi = {10.1103/PhysRevLett.134.106502},
  url = {https://link.aps.org/doi/10.1103/PhysRevLett.134.106502}
}

@article{wan2023topological,
  title={Topological exact flat bands in two-dimensional materials under periodic strain},
  author={Wan, Xiaohan and Sarkar, Siddhartha and Lin, Shi-Zeng and Sun, Kai},
  journal={Phys. Rev. Lett.},
  volume={130},
  number={21},
  pages={216401},
  year={2023},
  publisher={APS}
}

@article{gao2023untwisting,
  title = {Untwisting Moir\'e Physics: Almost Ideal Bands and Fractional {Chern} Insulators in Periodically Strained Monolayer Graphene},
  author = {Gao, Qiang and Dong, Junkai and Ledwith, Patrick and Parker, Daniel and Khalaf, Eslam},
  journal = {Phys. Rev. Lett.},
  volume = {131},
  issue = {9},
  pages = {096401},
  numpages = {9},
  year = {2023},
  month = {Aug},
  publisher = {American Physical Society},
  doi = {10.1103/PhysRevLett.131.096401},
  url = {https://link.aps.org/doi/10.1103/PhysRevLett.131.096401}
}

@article{wang2025closed,
  title = {Closed Band-Projected Density Algebra Must Be Girvin-MacDonald-Platzman},
  author = {Wang, Ziwei and Simon, Steven H.},
  journal = {Phys. Rev. Lett.},
  volume = {134},
  issue = {13},
  pages = {136502},
  numpages = {7},
  year = {2025},
  month = {Mar},
  publisher = {American Physical Society},
  doi = {10.1103/PhysRevLett.134.136502},
  url = {https://link.aps.org/doi/10.1103/PhysRevLett.134.136502}
}

@article{ledwith2023vortexability,
  title = {Vortexability: A unifying criterion for ideal fractional Chern insulators},
  author = {Ledwith, Patrick J. and Vishwanath, Ashvin and Parker, Daniel E.},
  journal = {Phys. Rev. B},
  volume = {108},
  issue = {20},
  pages = {205144},
  numpages = {20},
  year = {2023},
  month = {Nov},
  publisher = {American Physical Society},
  doi = {10.1103/PhysRevB.108.205144},
  url = {https://link.aps.org/doi/10.1103/PhysRevB.108.205144}
}

@article{zerofieldCFL23,
  title = {Composite Fermi Liquid at Zero Magnetic Field in Twisted ${\mathrm{MoTe}}_{2}$},
  author = {Dong, Junkai and Wang, Jie and Ledwith, Patrick J. and Vishwanath, Ashvin and Parker, Daniel E.},
  journal = {Phys. Rev. Lett.},
  volume = {131},
  issue = {13},
  pages = {136502},
  numpages = {10},
  year = {2023},
  month = {Sep},
  publisher = {American Physical Society},
  doi = {10.1103/PhysRevLett.131.136502},
  url = {https://link.aps.org/doi/10.1103/PhysRevLett.131.136502}
}

@article{Dong_decomposition23,
  title = {Many-body ground states from decomposition of ideal higher Chern bands: Applications to chirally twisted graphene multilayers},
  author = {Dong, Junkai and Ledwith, Patrick J. and Khalaf, Eslam and Lee, Jong Yeon and Vishwanath, Ashvin},
  journal = {Phys. Rev. Res.},
  volume = {5},
  issue = {2},
  pages = {023166},
  numpages = {17},
  year = {2023},
  month = {Jun},
  publisher = {American Physical Society},
  doi = {10.1103/PhysRevResearch.5.023166},
  url = {https://link.aps.org/doi/10.1103/PhysRevResearch.5.023166}
}

@article{haldane1983fractional,
  title = {Fractional Quantization of the Hall Effect: A Hierarchy of Incompressible Quantum Fluid States},
  author = {Haldane, F. D. M.},
  journal = {Phys. Rev. Lett.},
  volume = {51},
  issue = {7},
  pages = {605--608},
  numpages = {0},
  year = {1983},
  month = {Aug},
  publisher = {American Physical Society},
  doi = {10.1103/PhysRevLett.51.605},
  url = {https://link.aps.org/doi/10.1103/PhysRevLett.51.605}
}

@article{laughlin1983anomalous,
  title={Anomalous quantum Hall effect: An incompressible quantum fluid with fractionally charged excitations},
  author={Laughlin, Robert B},
  journal={Phys. Rev. Lett.},
  volume={50},
  number={18},
  pages={1395},
  year={1983},
  publisher={APS}
}

@article{haldane2018modular,
  title={A modular-invariant modified Weierstrass sigma-function as a building block for lowest-Landau-level wavefunctions on the torus},
  author={Haldane, FDM},
  journal={Journal of Mathematical Physics},
  volume={59},
  number={7},
  year={2018},
  publisher={AIP Publishing}
}

@article{shi2025effects,
  title={Effects of {Berry} curvature on ideal fractional {Chern} insulator many-body gaps},
  author={Shi, Jingtian and Cano, Jennifer and Morales-Dur{\'a}n, Nicol{\'a}s},
  journal={Preprint arXiv:2503.15900},
  year={2025}
}

@article{kourtis2018symmetry,
  title={Symmetry breaking and the fermionic fractional {Chern} insulator in topologically trivial bands},
  author={Kourtis, Stefanos},
  journal={Phys. Rev. B},
  volume={97},
  number={8},
  pages={085108},
  year={2018},
  publisher={APS}
}

@article{liu2025topological,
  title={Topological Order Without Band Topology in Moir\'e Graphene},
  author={Liu, Hui and Perea-Causin, Raul and Liu, Zhao and Bergholtz, Emil J},
  journal={Preprint arXiv:2510.15027},
  year={2025}
}

@article{bauer2016quantum,
  title={Quantum geometry and stability of the fractional quantum {Hall} effect in the {Hofstadter} model},
  author={Bauer, David and Jackson, TS and Roy, Rahul},
  journal={Phys. Rev. B},
  volume={93},
  number={23},
  pages={235133},
  year={2016},
  publisher={APS}
}

@article{wu2024quantum,
  title={Quantum-metric-induced quantum {Hall} conductance inversion and reentrant transition in fractional {Chern} insulators},
  author={Wu, Ang-Kun and Sarkar, Siddhartha and Wan, Xiaohan and Sun, Kai and Lin, Shi-Zeng},
  journal={Phys. Rev. Res.},
  volume={6},
  number={3},
  pages={L032063},
  year={2024},
  publisher={APS}
}

@article{parker2021field,
  title={Field-tuned and zero-field fractional {Chern} insulators in magic angle graphene},
  author={Parker, Daniel and Ledwith, Patrick and Khalaf, Eslam and Soejima, Tomohiro and Hauschild, Johannes and Xie, Yonglong and Pierce, Andrew and Zaletel, Michael P and Yacoby, Amir and Vishwanath, Ashvin},
  journal={Preprint arXiv:2112.13837},
  year={2021}
}

@article{aharonov1979ground,
  title = {Ground state of a spin-\textonehalf{} charged particle in a two-dimensional magnetic field},
  author = {Aharonov, Y. and Casher, A.},
  journal = {Phys. Rev. A},
  volume = {19},
  issue = {6},
  pages = {2461--2462},
  numpages = {0},
  year = {1979},
  month = {Jun},
  publisher = {American Physical Society},
  doi = {10.1103/PhysRevA.19.2461},
  url = {https://link.aps.org/doi/10.1103/PhysRevA.19.2461}
}

@misc{ArovasQHE,
  author       = {Arovas, Daniel},
  title        = {Lecture Notes on Quantum Hall Effect
(A Work in Progress)},
  year         = {2024},
  howpublished = {\url{https://courses.physics.ucsd.edu/2019/Spring/physics230/LECTURES/QHE.pdf}},
  url          = {https://courses.physics.ucsd.edu/2019/Spring/physics230/LECTURES/QHE.pdf}
}

@article{trugman1985exact,
  title={Exact results for the fractional quantum Hall effect with general interactions},
  author={Trugman, Stuart Alan and Kivelson, S},
  journal={Phys. Rev. B},
  volume={31},
  number={8},
  pages={5280},
  year={1985},
  publisher={APS}
}

@article{macdonald1994introduction,
  title={Introduction to the physics of the quantum Hall regime},
  author={Macdonald, Allan H},
  journal={Preprint cond-mat/9410047},
  year={1994},
  url={https://arxiv.org/abs/cond-mat/9410047}
}

@article{tong2016lectures,
  title={Lectures on the quantum Hall effect},
  author={Tong, David},
  journal={Preprint arXiv:1606.06687},
  year={2016}
}

@article{girvin1999quantumhalleffectnovel,
      title={The Quantum Hall Effect: Novel Excitations and Broken Symmetries}, 
      author={Steven M. Girvin},
      year={1999},
      journal={Preprint cond-mat/9907002},
      url={https://arxiv.org/abs/cond-mat/9907002}
}

@article{CAN2015752,
title = {Geometry of quantum Hall states: Gravitational anomaly and transport coefficients},
journal = {Annals of Physics},
volume = {362},
pages = {752-794},
year = {2015},
issn = {0003-4916},
doi = {https://doi.org/10.1016/j.aop.2015.02.013},
url = {https://www.sciencedirect.com/science/article/pii/S0003491615000482},
author = {Tankut Can and Michael Laskin and Paul B. Wiegmann}
}

@article{peres2006electronic,
  title={Electronic properties of disordered two-dimensional carbon},
  author={Peres, NMR and Guinea, F and Castro Neto, AH},
  journal={Phys. Rev. B},
  volume={73},
  number={12},
  pages={125411},
  year={2006},
  publisher={APS}
}

@article{gusynin2005unconventional,
  title={Unconventional integer quantum Hall effect in graphene},
  author={Gusynin, VP and Sharapov, SG},
  journal={Phys. Rev. Lett.},
  volume={95},
  number={14},
  pages={146801},
  year={2005},
  publisher={APS}
}

@article{novoselov2005two,
  title={Two-dimensional gas of massless Dirac fermions in graphene},
  author={Novoselov, Kostya S and Geim, Andre K and Morozov, Sergei Vladimirovich and Jiang, Dingde and Katsnelson, Michail I and Grigorieva, Irina V and Dubonos, Sergey V and Firsov, Alexandr A},
  journal={Nature},
  volume={438},
  number={7065},
  pages={197--200},
  year={2005},
  publisher={Nature Publishing Group UK London}
}

@article{zhang2005experimental,
  title={Experimental observation of the quantum Hall effect and Berry's phase in graphene},
  author={Zhang, Yuanbo and Tan, Yan-Wen and Stormer, Horst L and Kim, Philip},
  journal={Nature},
  volume={438},
  number={7065},
  pages={201--204},
  year={2005},
  publisher={Nature Publishing Group UK London}
}

@article{xie2021fractional,
  title={Fractional Chern insulators in magic-angle twisted bilayer graphene},
  author={Xie, Yonglong and Pierce, Andrew T. and Park, Jeong Min and Parker, Daniel E. and Khalaf, Eslam and Ledwith, Patrick and Cao, Yuan and Lee, Seung Hwan and Chen, Shaowen and Forrester, Patrick R. and Watanabe, Kenji and Taniguchi, Takashi and Jarillo-Herrero, Pablo and Vishwanath, Ashvin},
  journal={Nature},
  volume={600},
  number={7889},
  pages={439--443},
  year={2021},
  publisher={Nature Publishing Group UK London}
}

@article{cai2023signatures,
  title={Signatures of fractional quantum anomalous Hall states in twisted MoTe2},
  author={Cai, Jiaqi and Anderson, Eric and Wang, Chong and Zhang, Xiaowei and Liu, Xiaoyu and Holtzmann, William and Zhang, Yinong and Fan, Fengren and Taniguchi, Takashi and Watanabe, Kenji and others},
  journal={Nature},
  volume={622},
  number={7981},
  pages={63--68},
  year={2023},
  publisher={Nature Publishing Group UK London}
}

@article{zeng2023thermodynamic,
  title={Thermodynamic evidence of fractional Chern insulator in moir{\'e} MoTe2},
  author={Zeng, Yihang and Xia, Zhengchao and Kang, Kaifei and Zhu, Jiacheng and Kn{\"u}ppel, Patrick and Vaswani, Chirag and Watanabe, Kenji and Taniguchi, Takashi and Mak, Kin Fai and Shan, Jie},
  journal={Nature},
  volume={622},
  number={7981},
  pages={69--73},
  year={2023},
  publisher={Nature Publishing Group UK London}
}

@article{park2023observation,
  title={Observation of fractionally quantized anomalous Hall effect},
  author={Park, Heonjoon and Cai, Jiaqi and Anderson, Eric and Zhang, Yinong and Zhu, Jiayi and Liu, Xiaoyu and Wang, Chong and Holtzmann, William and Hu, Chaowei and Liu, Zhaoyu and others},
  journal={Nature},
  volume={622},
  number={7981},
  pages={74--79},
  year={2023},
  publisher={Nature Publishing Group UK London}
}

@article{xu2023observation,
  title={Observation of integer and fractional quantum anomalous Hall effects in twisted bilayer MoTe 2},
  author={Xu, Fan and Sun, Zheng and Jia, Tongtong and Liu, Chang and Xu, Cheng and Li, Chushan and Gu, Yu and Watanabe, Kenji and Taniguchi, Takashi and Tong, Bingbing and others},
  journal={Phys. Rev. X},
  volume={13},
  number={3},
  pages={031037},
  year={2023},
  publisher={APS}
}

@article{wu2019topological,
  title = {Topological Insulators in Twisted Transition Metal Dichalcogenide Homobilayers},
  author = {Wu, Fengcheng and Lovorn, Timothy and Tutuc, Emanuel and Martin, Ivar and MacDonald, A. H.},
  journal = {Phys. Rev. Lett.},
  volume = {122},
  issue = {8},
  pages = {086402},
  numpages = {5},
  year = {2019},
  month = {Feb},
  publisher = {American Physical Society},
  doi = {10.1103/PhysRevLett.122.086402},
  url = {https://link.aps.org/doi/10.1103/PhysRevLett.122.086402}
}

@article{FCWu_VariatoinalMapping25,
  title = {Variational mapping of Chern bands to Landau levels: Application to fractional Chern insulators in twisted ${\mathrm{MoTe}}_{2}$},
  author = {Li, Bohao and Wu, Fengcheng},
  journal = {Phys. Rev. B},
  volume = {111},
  issue = {12},
  pages = {125122},
  numpages = {12},
  year = {2025},
  month = {Mar},
  publisher = {American Physical Society},
  doi = {10.1103/PhysRevB.111.125122},
  url = {https://link.aps.org/doi/10.1103/PhysRevB.111.125122}
}

@ARTICLE{FCWu_GLL26,
       author = {{Li}, Bohao and {Ouyang}, Yunze and {Wu}, Fengcheng},
        title = "{Abelian and non-Abelian fractionalized states in twisted MoTe$_2$: A generalized Landau-level theory}",
      journal = {arXiv e-prints},
         year = 2026,
        month = jan,
          eid = {arXiv:2601.13169},
        pages = {arXiv:2601.13169},
          doi = {10.48550/arXiv.2601.13169},
archivePrefix = {arXiv},
       eprint = {2601.13169},
 primaryClass = {cond-mat.mes-hall},
       adsurl = {https://ui.adsabs.harvard.edu/abs/2026arXiv260113169L}
}

@article{devakul2021magic,
  title={Magic in twisted transition metal dichalcogenide bilayers},
  author={Devakul, Trithep and Cr{\'e}pel, Valentin and Zhang, Yang and Fu, Liang},
  journal={Nat. Comm.},
  volume={12},
  number={1},
  pages={6730},
  year={2021},
  publisher={Nature Publishing Group UK London}
}

@article{li2021spontaneous,
  title={Spontaneous fractional Chern insulators in transition metal dichalcogenide moir{\'e} superlattices},
  author={Li, Heqiu and Kumar, Umesh and Sun, Kai and Lin, Shi-Zeng},
  journal={Phys. Rev. Res.},
  volume={3},
  number={3},
  pages={L032070},
  year={2021},
  publisher={APS}
}

@article{crepel2023anomalous,
  title={Anomalous Hall metal and fractional Chern insulator in twisted transition metal dichalcogenides},
  author={Cr{\'e}pel, Valentin and Fu, Liang},
  journal={Phys. Rev. B},
  volume={107},
  number={20},
  pages={L201109},
  year={2023},
  publisher={APS}
}

@article{jia2024moire,
  title={Moir{\'e} fractional {Chern} insulators. {I}. First-principles calculations and continuum models of twisted bilayer {MoTe$_2$}},
  author={Jia, Yujin and Yu, Jiabin and Liu, Jiaxuan and Herzog-Arbeitman, Jonah and Qi, Ziyue and Pi, Hanqi and Regnault, Nicolas and Weng, Hongming and Bernevig, B Andrei and Wu, Quansheng},
  journal={Phys. Rev. B},
  volume={109},
  number={20},
  pages={205121},
  year={2024},
  publisher={APS}
}

@article{wang2025higher,
  title = {Higher {Landau}-Level Analogs and Signatures of Non-{Abelian} States in Twisted Bilayer ${\mathrm{MoTe}}_{2}$},
  author = {Wang, Chong and Zhang, Xiao-Wei and Liu, Xiaoyu and Wang, Jie and Cao, Ting and Xiao, Di},
  journal = {Phys. Rev. Lett.},
  volume = {134},
  issue = {7},
  pages = {076503},
  numpages = {7},
  year = {2025},
  month = {Feb},
  publisher = {American Physical Society},
  doi = {10.1103/PhysRevLett.134.076503},
  url = {https://link.aps.org/doi/10.1103/PhysRevLett.134.076503}
}

@article{ghorashi2023topological,
  title={Topological and stacked flat bands in bilayer graphene with a superlattice potential},
  author={Ghorashi, Sayed Ali Akbar and Dunbrack, Aaron and Abouelkomsan, Ahmed and Sun, Jiacheng and Du, Xu and Cano, Jennifer},
  journal={Phys. Rev. Lett.},
  volume={130},
  number={19},
  pages={196201},
  year={2023},
  publisher={APS}
}

@article{ault2026optimizing,
  title={Optimizing superlattice bilayer graphene for a fractional Chern insulator},
  author={Ault-McCoy, Dathan and Lhachemi, M Nabil Y and Dunbrack, Aaron and Ghorashi, Sayed Ali Akbar and Cano, Jennifer},
  journal={Phys. Rev. B},
  volume={113},
  number={7},
  pages={075140},
  year={2026},
  publisher={APS}
}

@article{zeng2024gate,
  title={Gate-tunable topological phases in superlattice modulated bilayer graphene},
  author={Zeng, Yongxin and Wolf, Tobias MR and Huang, Chunli and Wei, Nemin and Ghorashi, Sayed Ali Akbar and MacDonald, Allan H and Cano, Jennifer},
  journal={Phys. Rev. B},
  volume={109},
  number={19},
  pages={195406},
  year={2024},
  publisher={APS}
}

@article{lu2024fractional,
  title={Fractional quantum anomalous {Hall} effect in multilayer graphene},
  author={Lu, Zhengguang and Han, Tonghang and Yao, Yuxuan and Reddy, Aidan P and Yang, Jixiang and Seo, Junseok and Watanabe, Kenji and Taniguchi, Takashi and Fu, Liang and Ju, Long},
  journal={Nature},
  volume={626},
  number={8000},
  pages={759--764},
  year={2024},
  publisher={Nature Publishing Group UK London}
}

@article{xie2025tunable,
  title={Tunable fractional Chern insulators in rhombohedral graphene superlattices},
  author={Xie, Jian and Huo, Zihao and Lu, Xin and Feng, Zuo and Zhang, Zaizhe and Wang, Wenxuan and Yang, Qiu and Watanabe, Kenji and Taniguchi, Takashi and Liu, Kaihui and others},
  journal={Nature Materials},
  volume={24},
  number={7},
  pages={1042--1048},
  year={2025},
  publisher={Nature Publishing Group UK London}
}

@article{tan2025ideal,
  title={The ideal limit of rhombohedral graphene: Interaction-induced layer-skyrmion lattices and their collective excitations},
  author={Tan, Tixuan and Ledwith, Patrick J and Devakul, Trithep},
  journal={Preprint arXiv:2511.07402},
  year={2025}
}

@article{bernevig2025berry,
  title={" Berry Trashcan" Model of Interacting Electrons in Rhombohedral Graphene},
  author={Bernevig, B Andrei and Kwan, Yves H},
  journal={Preprint arXiv:2503.09692},
  year={2025}
}

@article{estienne2023ideal,
  title={Ideal Chern bands as Landau levels in curved space},
  author={Estienne, Benoit and Regnault, Nicolas and Cr{\'e}pel, Valentin},
  journal={Phys. Rev. Res.},
  volume={5},
  number={3},
  pages={L032048},
  year={2023},
  publisher={APS}
}

@article{KapitMuellerModel,
  title = {Exact Parent Hamiltonian for the Quantum Hall States in a Lattice},
  author = {Kapit, Eliot and Mueller, Erich},
  journal = {Phys. Rev. Lett.},
  volume = {105},
  issue = {21},
  pages = {215303},
  numpages = {4},
  year = {2010},
  month = {Nov},
  publisher = {American Physical Society},
  doi = {10.1103/PhysRevLett.105.215303},
  url = {https://link.aps.org/doi/10.1103/PhysRevLett.105.215303}
}

@article{Xiao_2010,
  title = {Berry phase effects on electronic properties},
  author = {Xiao, Di and Chang, Ming-Che and Niu, Qian},
  journal = {Rev. Mod. Phys.},
  volume = {82},
  issue = {3},
  pages = {1959--2007},
  numpages = {0},
  year = {2010},
  month = {Jul},
  publisher = {American Physical Society},
  doi = {10.1103/RevModPhys.82.1959},
  url = {https://link.aps.org/doi/10.1103/RevModPhys.82.1959}
}

@article{FCI_Bernevig_PRX,
  title = {Fractional Chern Insulator},
  author = {Regnault, N. and Bernevig, B. Andrei},
  journal = {Phys. Rev. X},
  volume = {1},
  issue = {2},
  pages = {021014},
  numpages = {14},
  year = {2011},
  month = {Dec},
  publisher = {American Physical Society},
  doi = {10.1103/PhysRevX.1.021014},
  url = {https://link.aps.org/doi/10.1103/PhysRevX.1.021014}
}

@article{Review_moire_FCI,
   author = "Cao, Ting and Fu, Liang and Ju, Long and Xiao, Di and Xu, Xiaodong",
   title = "Fractional Quantum Anomalous Hall Effect", 
   journal= "Annu. Rev. of Condens. Matter. Phys.",
   year = "2026",
   volume = "17",
   number = "Volume 17, 2026",
   pages = "233-256",
   doi = "https://doi.org/10.1146/annurev-conmatphys-031524-071133",
   url = "https://www.annualreviews.org/content/journals/10.1146/annurev-conmatphys-031524-071133",
   publisher = "Annual Reviews",
   issn = "1947-5462",
   type = "Journal Article",
  }

@article{Torma_Essay,
  title = {Essay: Where Can Quantum Geometry Lead Us?},
  author = {T\"orm\"a, P\"aivi},
  journal = {Phys. Rev. Lett.},
  volume = {131},
  issue = {24},
  pages = {240001},
  numpages = {7},
  year = {2023},
  month = {Dec},
  publisher = {American Physical Society},
  doi = {10.1103/PhysRevLett.131.240001},
  url = {https://link.aps.org/doi/10.1103/PhysRevLett.131.240001}
}

@ARTICLE{ZhaoReview,
       author = {{Liu}, Zhao and {Bergholtz}, Emil J.},
        title = "{Recent Developments in Fractional Chern Insulators}",
      journal = {arXiv e-prints},
         year = 2022,
        month = aug,
          eid = {arXiv:2208.08449},
        pages = {arXiv:2208.08449},
          doi = {10.48550/arXiv.2208.08449},
archivePrefix = {arXiv},
       eprint = {2208.08449},
 primaryClass = {cond-mat.mes-hall},
       adsurl = {https://ui.adsabs.harvard.edu/abs/2022arXiv220808449L}
}

@article{HaldaneGeometryFQH,
  title = {Geometrical Description of the Fractional Quantum Hall Effect},
  author = {Haldane, F. D. M.},
  journal = {Phys. Rev. Lett.},
  volume = {107},
  issue = {11},
  pages = {116801},
  numpages = {5},
  year = {2011},
  month = {Sep},
  publisher = {American Physical Society},
  doi = {10.1103/PhysRevLett.107.116801},
  url = {https://link.aps.org/doi/10.1103/PhysRevLett.107.116801}
}

@article{haldanetorus2,
	Author = {Haldane, F. D. M.},
	Doi = {10.1103/PhysRevLett.55.2095},
	Issue = {20},
	Journal = {Phys. Rev. Lett.},
	Month = {Nov},
	Numpages = {0},
	Pages = {2095--2098},
	Publisher = {American Physical Society},
	Title = {Many-Particle Translational Symmetries of Two-Dimensional Electrons at Rational Landau-Level Filling},
	Url = {https://link.aps.org/doi/10.1103/PhysRevLett.55.2095},
	Volume = {55},
	Year = {1985}}

@article{haldanetorus1,
	Author = {Haldane, F. D. M. and Rezayi, E. H.},
	Doi = {10.1103/PhysRevB.31.2529},
	Issue = {4},
	Journal = {Phys. Rev. B},
	Month = {Feb},
	Numpages = {0},
	Pages = {2529--2531},
	Publisher = {American Physical Society},
	Title = {Periodic Laughlin-Jastrow wave functions for the fractional quantized Hall effect},
	Url = {https://link.aps.org/doi/10.1103/PhysRevB.31.2529},
	Volume = {31},
	Year = {1985}}

@article{haldanemodularinv,
	Author = {Haldane,F. D. M.},
	Doi = {10.1063/1.5042618},
	Eprint = {https://doi.org/10.1063/1.5042618},
	Journal = {Journal of Mathematical Physics},
	Number = {7},
	Pages = {071901},
	Title = {A modular-invariant modified Weierstrass sigma-function as a building block for lowest-Landau-level wavefunctions on the torus},
	Url = {https://doi.org/10.1063/1.5042618},
	Volume = {59},
	Year = {2018}}

@article{JW_MonteCarlo,
  title = {Lattice Monte Carlo for quantum Hall states on a torus},
  author = {Wang, Jie and Geraedts, Scott D. and Rezayi, E. H. and Haldane, F. D. M.},
  journal = {Phys. Rev. B},
  volume = {99},
  issue = {12},
  pages = {125123},
  numpages = {13},
  year = {2019},
  month = {Mar},
  publisher = {American Physical Society},
  doi = {10.1103/PhysRevB.99.125123},
  url = {https://link.aps.org/doi/10.1103/PhysRevB.99.125123}
}

@article{JW_CTBG_Nodal,
  title = {Chiral approximation to twisted bilayer graphene: Exact intravalley inversion symmetry, nodal structure, and implications for higher magic angles},
  author = {Wang, Jie and Zheng, Yunqin and Millis, Andrew J. and Cano, Jennifer},
  journal = {Phys. Rev. Res.},
  volume = {3},
  issue = {2},
  pages = {023155},
  numpages = {17},
  year = {2021},
  month = {May},
  publisher = {American Physical Society},
  doi = {10.1103/PhysRevResearch.3.023155},
  url = {https://link.aps.org/doi/10.1103/PhysRevResearch.3.023155}
}

@ARTICLE{JieLiangFuDirac22,
       author = {{Dong}, Junkai and {Wang}, Jie and {Fu}, Liang},
        title = "{Dirac electron under periodic magnetic field: Platform for fractional Chern insulator and generalized Wigner crystal}",
      journal = {arXiv e-prints},
         year = 2022,
        month = aug,
          eid = {arXiv:2208.10516},
        pages = {arXiv:2208.10516},
          doi = {10.48550/arXiv.2208.10516},
archivePrefix = {arXiv},
       eprint = {2208.10516},
 primaryClass = {cond-mat.mes-hall},
       adsurl = {https://ui.adsabs.harvard.edu/abs/2022arXiv220810516D}
}

@article{Shi_ACband_tTMD24,
  title = {Adiabatic approximation and Aharonov-Casher bands in twisted homobilayer transition metal dichalcogenides},
  author = {Shi, Jingtian and Morales-Dur\'an, Nicol\'as and Khalaf, Eslam and MacDonald, A. H.},
  journal = {Phys. Rev. B},
  volume = {110},
  issue = {3},
  pages = {035130},
  numpages = {17},
  year = {2024},
  month = {Jul},
  publisher = {American Physical Society},
  doi = {10.1103/PhysRevB.110.035130},
  url = {https://link.aps.org/doi/10.1103/PhysRevB.110.035130}
}

@article{gao2025quantum,
  title={Quantum geometry phenomena in condensed matter systems},
  author={Gao, Anyuan and Nagaosa, Naoto and Ni, Ni and Xu, Su-Yang},
  journal={Preprint arXiv:2508.00469},
  year={2025}
}

@article{jiang2025revealing,
  title={Revealing quantum geometry in nonlinear quantum materials},
  author={Jiang, Yiyang and Holder, Tobias and Yan, Binghai},
  journal={Reports on Progress in Physics},
  volume={88},
  number={7},
  pages={076502},
  year={2025},
  publisher={IOP Publishing}
}

@article{andrei2021marvels,
  title={The marvels of moir{\'e} materials},
  author={Andrei, Eva Y and Efetov, Dmitri K and Jarillo-Herrero, Pablo and MacDonald, Allan H and Mak, Kin Fai and Senthil, T and Tutuc, Emanuel and Yazdani, Ali and Young, Andrea F},
  journal={Nat. Rev. Mater.},
  volume={6},
  number={3},
  pages={201--206},
  year={2021},
  publisher={Nature Publishing Group UK London}
}

@article{nuckolls2024microscopic,
  title={A microscopic perspective on moir{\'e} materials},
  author={Nuckolls, Kevin P and Yazdani, Ali},
  journal={Nat. Rev. Mater.},
  volume={9},
  number={7},
  pages={460--480},
  year={2024},
  publisher={Nature Publishing Group UK London}
}

@article{sheng2011fractional,
  title={Fractional quantum Hall effect in the absence of Landau levels},
  author={Sheng, DN and Gu, Zheng-Cheng and Sun, Kai and Sheng, L},
  journal={Nat. Comm.},
  volume={2},
  number={1},
  pages={389},
  year={2011},
  publisher={Nature Publishing Group UK London}
}

@article{neupert2011fractional,
  title={Fractional quantum Hall states at zero magnetic field},
  author={Neupert, Titus and Santos, Luiz and Chamon, Claudio and Mudry, Christopher},
  journal={Phys. Rev. Lett.},
  volume={106},
  number={23},
  pages={236804},
  year={2011},
  publisher={APS}
}

@article{QianNiuSemiClassical99,
  title = {Wave-packet dynamics in slowly perturbed crystals: Gradient corrections and Berry-phase effects},
  author = {Sundaram, Ganesh and Niu, Qian},
  journal = {Phys. Rev. B},
  volume = {59},
  issue = {23},
  pages = {14915--14925},
  numpages = {0},
  year = {1999},
  month = {Jun},
  publisher = {American Physical Society},
  doi = {10.1103/PhysRevB.59.14915},
  url = {https://link.aps.org/doi/10.1103/PhysRevB.59.14915}
}

@article{Haldane_anomalousHall04,
  title = {Berry Curvature on the Fermi Surface: Anomalous Hall Effect as a Topological Fermi-Liquid Property},
  author = {Haldane, F. D. M.},
  journal = {Phys. Rev. Lett.},
  volume = {93},
  issue = {20},
  pages = {206602},
  numpages = {4},
  year = {2004},
  month = {Nov},
  publisher = {American Physical Society},
  doi = {10.1103/PhysRevLett.93.206602},
  url = {https://link.aps.org/doi/10.1103/PhysRevLett.93.206602}
}

@article{RMP_Resta_Polarization94,
  title = {Macroscopic polarization in crystalline dielectrics: the geometric phase approach},
  author = {Resta, Raffaele},
  journal = {Rev. Mod. Phys.},
  volume = {66},
  issue = {3},
  pages = {899--915},
  numpages = {0},
  year = {1994},
  month = {Jul},
  publisher = {American Physical Society},
  doi = {10.1103/RevModPhys.66.899},
  url = {https://link.aps.org/doi/10.1103/RevModPhys.66.899}
}

@article{Vanderbilt_Polarization93,
  title = {Theory of polarization of crystalline solids},
  author = {King-Smith, R. D. and Vanderbilt, David},
  journal = {Phys. Rev. B},
  volume = {47},
  issue = {3},
  pages = {1651--1654},
  numpages = {0},
  year = {1993},
  month = {Jan},
  publisher = {American Physical Society},
  doi = {10.1103/PhysRevB.47.1651},
  url = {https://link.aps.org/doi/10.1103/PhysRevB.47.1651}
}

@article{RMP_Vanderbilt84,
  title = {Maximally localized Wannier functions: Theory and applications},
  author = {Marzari, Nicola and Mostofi, Arash A. and Yates, Jonathan R. and Souza, Ivo and Vanderbilt, David},
  journal = {Rev. Mod. Phys.},
  volume = {84},
  issue = {4},
  pages = {1419--1475},
  numpages = {0},
  year = {2012},
  month = {Oct},
  publisher = {American Physical Society},
  doi = {10.1103/RevModPhys.84.1419},
  url = {https://link.aps.org/doi/10.1103/RevModPhys.84.1419}
}

@article{Niu_Quantization85,
  title = {Quantized Hall conductance as a topological invariant},
  author = {Niu, Qian and Thouless, D. J. and Wu, Yong-Shi},
  journal = {Phys. Rev. B},
  volume = {31},
  issue = {6},
  pages = {3372--3377},
  numpages = {0},
  year = {1985},
  month = {Mar},
  publisher = {American Physical Society},
  doi = {10.1103/PhysRevB.31.3372},
  url = {https://link.aps.org/doi/10.1103/PhysRevB.31.3372}
}

@article{Du_Graviton,
	author = {Liang, Jiehui and Liu, Ziyu and Yang, Zihao and Huang, Yuelei and Wurstbauer, Ursula and Dean, Cory R. and West, Ken W. and Pfeiffer, Loren N. and Du, Lingjie and Pinczuk, Aron},
	date = {2024/04/01},
	doi = {10.1038/s41586-024-07201-w},
	id = {Liang2024},
	isbn = {1476-4687},
	journal = {Nature},
	number = {8006},
	pages = {78--83},
	title = {Evidence for chiral graviton modes in fractional quantum Hall liquids},
	url = {https://doi.org/10.1038/s41586-024-07201-w},
	volume = {628},
	year = {2024}}

@article{SWM_SumRule,
  title = {Polarization and localization in insulators: Generating function approach},
  author = {Souza, Ivo and Wilkens, Tim and Martin, Richard M.},
  journal = {Phys. Rev. B},
  volume = {62},
  issue = {3},
  pages = {1666--1683},
  numpages = {0},
  year = {2000},
  month = {Jul},
  publisher = {American Physical Society},
  doi = {10.1103/PhysRevB.62.1666},
  url = {https://link.aps.org/doi/10.1103/PhysRevB.62.1666}
}

\newpage

\section{SUPPLEMENTAL MATERIAL}

\subsection{Geometric bound and null vectors}
\label{sec:QGTproofs}

In this section we derive the bounds on the QGT and their relation to its null vectors.

The QGT is defined as
\begin{equation}
    Q^{ab}_{\bm k} = g^{ab}_{\bm k} + \frac{i\epsilon^{ab}}{2} \Omega_{\bm k}.
\end{equation}
Since it is Hermitian and non-negative, it has two real and non-negative eigenvalues, $\lambda_{1,2}(\bm k)$, and its determinant must be non-negative. This yields:
\begin{eqnarray}
    0 &\leq& 2\det Q_{\bm k} = \epsilon_{ac} \epsilon_{bd} \left(g^{ab}_{\bm k} + \frac{i\epsilon^{ab}}{2} \Omega_{\bm k}\right) \left( g^{cd}_{\bm k} + \frac{i\epsilon^{cd}}{2} \Omega_{\bm k}\right) \nonumber\\
    &=& \epsilon_{ac} \epsilon_{bd} g^{ab}_{\bm k} g^{cd}_{\bm k} - \frac{1}{4} \epsilon_{ac} \epsilon_{bd} \epsilon^{ab} \epsilon^{cd} |\Omega_{\bm k}|^2 = 2\det g_{\bm k} - \frac{1}{2} |\Omega_{\bm k}|^2,
\end{eqnarray}
which yields the determinant bound $\sqrt{\det g_{\bm k}} \geq |\Omega_{\bm k}|/2$. The inequality saturates if and only if one of $\lambda_{1,2}(\bm k)$ is zero; otherwise $\det Q_{\bm k}$ is strictly positive. 
Thus, saturation of the determinant bound implies that $Q^{ab}_{\bm{k}}$ has a null vector.
In general, the null vector can be $\bm{k}$-dependent -- in fact, every two-band model in 2D saturates the determinant bound at every $\bm{k}$ with a $\bm{k}$-dependent null vector~\cite{mera2021}.

Using the inequality of arithmetic and geometric means, the trace of a $2\times 2$ non-negative Hermitian matrix is lower bounded by twice the square root of its determinant, yielding
the following quantum geometric bound,
\begin{equation}
    {\rm Tr}_{h} g_{\bm k} \geq 2\sqrt{\det g_{\bm k}} \geq |\Omega_{\bm k}|,
    \label{eq:qgtbounds2}
\end{equation}
for any positive symmetric matrix $h_{ab}$ with $\det h = 1$.

We now show that saturation of both bounds with a $\bm{k}$-independent matrix $h$ -- i.e., satisfying the ideal band condition -- implies that the QGT has a $\bm{k}$-independent null vector.
To this end, we rewrite $h$ in terms of complex vectors $w,w^*$ as
\begin{equation}
    h_{ab} = w_a^* w_b + w_a w_b^*, \quad i\epsilon_{ab} = w_a^* w_b - w_a w_b^*.
\end{equation}
The non-negativity of the QGT implies
\begin{equation}
   0\leq  w_a^* Q^{ab}_{\bm k} w_b= \frac{1}{2} \left[ h_{ab} + i\epsilon_{ab} \right]\left( g^{ab}_{\bm k} + \frac{i\epsilon^{ab}}{2} \Omega_{\bm k} \right) = \frac{1}{2}\left( h_{ab} g^{ab}_{\bm{k}} -  \Omega_{\bm{k}} \right) .
   \label{sup_geo_bound}
\end{equation}
Saturation of both bounds in Eq.~\eqref{eq:qgtbounds2} implies that the right-hand-side of Eq.~\eqref{sup_geo_bound} vanishes, which then implies that $w$ is a ($\bm{k}$-independent) null vector of $Q_{\bm{k}}$.

Thus, the ideal band condition is much stronger than the determinant condition: it is a global trace bound that requires the null vector of the QGT be momentum-independent.

\subsection{Translation operator in the lowest Landau level}
\label{sec:translationLLL}

In this section, we derive how the translation operator acts on wave functions in the LLL.
Starting from the definition of the translation operator,
\begin{equation}
    t(\bm d) = \exp(i\bm d\times\bm R/\ell_B^2) = \exp\left[(d^*\hat{b}^\dag - d \hat{b})/\ell_B\right],
\end{equation}
we expand it using the Baker-Campbell-Hausdorff relation $e^{A+B} = e^{A} e^{B} e^{-\frac12[A,B]}$ as:
\begin{equation}
    t(\bm d) = e^{d^* \hat b^\dag - d \hat b} = e^{-\frac{1}{2}|d|^2} e^{d^* \hat b^\dag} e^{-d \hat b} = e^{-\frac{1}{2}|d|^2} e^{d^* (-\bar\partial + z/2)} e^{-d (\partial + \bar z/2)},
\end{equation}
where we have set $\ell_B = 1$ and in the last equality used the explicit form of the raising and lowering operators,
\begin{equation}
    \hat{b} = \partial_z  + \frac{\bar{z}}{2}, \quad \hat{b}^\dagger = - \partial_{\bar{z}}  + \frac{z}{2}.
\end{equation}

To act with this operator on a wave function, recall the more familiar translation operator $e^{d\partial_z} f(z) = f(z+d)$,
which can be derived quite generally using the Taylor expansion for a single-variable function,
\begin{equation}
    f(z+d) = f(z) + d \partial_z f(z) + \frac{1}{2!}d^2\partial_z^2 f(z) + \cdots = e^{d\partial_z} f(z) .
\end{equation}
Acting on a LLL wave function yields:
\begin{equation}
     e^{-d (\partial + \bar z/2)} f(z) e^{-\frac{1}{2} |z|^2} = f(z-d) e^{-\frac{1}{2} |z|^2}
\end{equation}
and ultimately
\begin{equation}
    t(\mathbf{d}) f(z) e^{-\frac{1}{2}|z|^2} = e^{-\frac{1}{2}|d|^2}e^{zd^*} f(z-d) e^{-\frac{1}{2} |z|^2} ,
\end{equation}
which can be rewritten as,
\begin{equation}
    t(\mathbf{d}) \psi_\text{LLL}(\mathbf{r}) = \psi_\text{LLL}(\mathbf{r} - \mathbf{d}) e^{\frac{1}{2} ( z d^* - z^* d ) }.
\end{equation}

\subsection{Kapit-Mueller model}
\label{sec:kapitmueller}

In this section, we briefly describe the Kapit-Mueller (KM) model, which is a lattice model with an ideal band. We then illustrate how changing the orbital embedding can change the band from ideal to vortexable. 

The KM model is a special type of Hofstadter model. Denoting the lattice vectors by $\bm a_{1,2}$, with unit cell area $|\bm a_1 \times \bm a_2| = 2\pi S$, each unit cell contains $\phi < 1$ flux. A general form of the Hofstadter Hamiltonian can be written as
\begin{equation}
    H_{\rm Hof} = \sum_{\bm d \in \Lambda} J(\bm d) t(\bm d),
\end{equation}
where $\Lambda = \{m\bm a_1+n\bm a_2\}$, $J(\bm d)$ is a hopping term, and $t(\bm d)$ is the magnetic translation operator. 
In the KM model, the hopping parameter exhibits Gaussian decay on the lattice, $J(\bm d) = \exp\left[-(1-\phi)\bm d^2 S\right]$.
In this special case, the spectrum is non-negative with an exact flat band whose wavefunction is a LLL wavefunction sampled on lattice sites, i.e., $H_{\rm KM}|\psi_{\bm k}\rangle = 0$, where the flat band wave function is given by
\begin{equation}
    |\psi_{\bm k} \rangle = \sum_{\bm r\in\Lambda} \Phi_{\bm k}(\bm r) |\bm r\rangle,
\end{equation}
where $\Phi_{\bm k}(\bm r)$ is the LLL wavefunction and $|\bm r\rangle$ represents a localized orthonormal basis of states.
As a consequence of its ideal band geometry, the KM model supports a Laughlin state of bosons at $\nu = 1/2$ as an exact ground state with on-site Hubbard repulsion~\cite{KapitMuellerModel}. 

\begin{figure}
    \includegraphics[width = \textwidth]{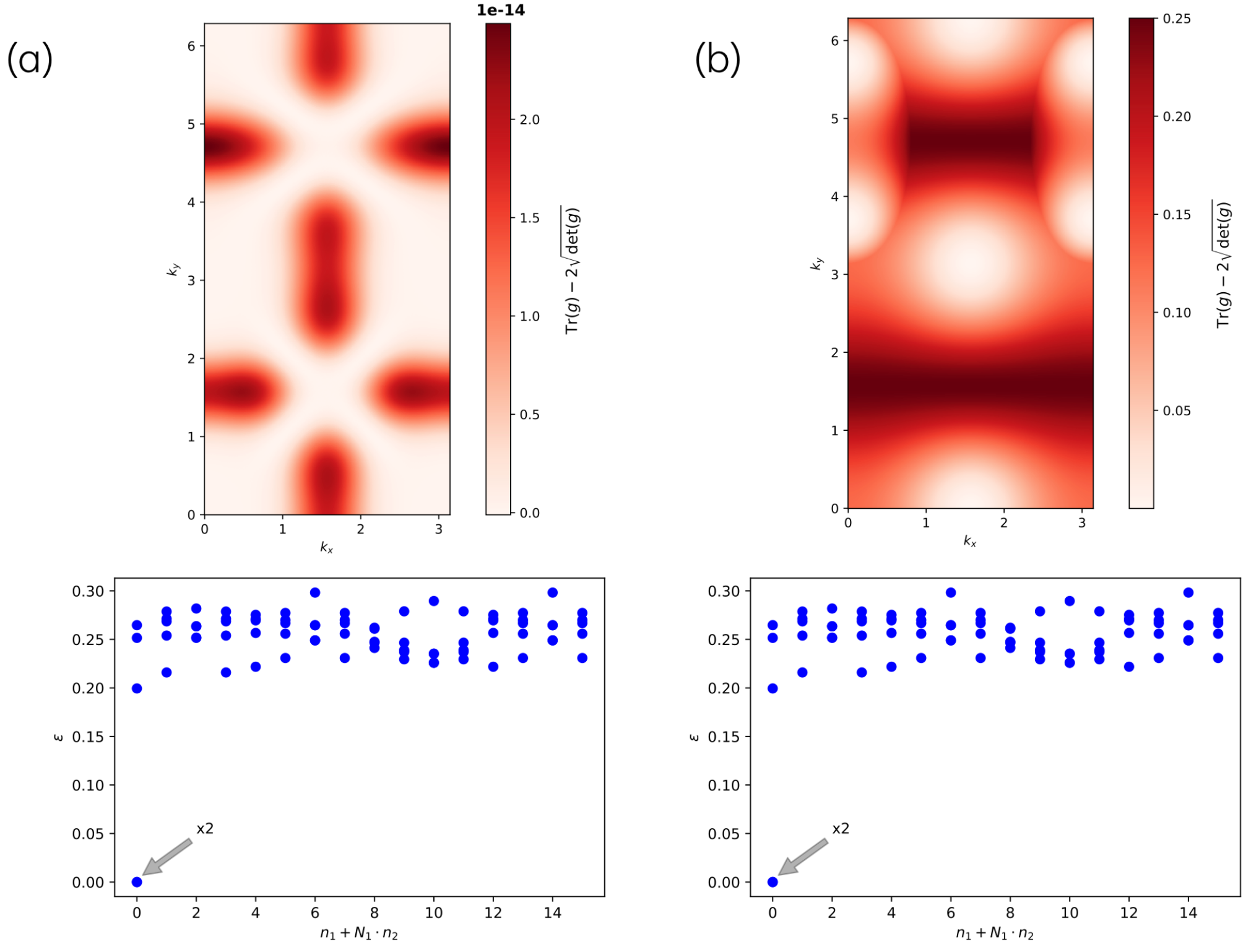}
    \caption{Deviation from the trace condition for the flat band in the KM model (top) and many-body spectrum with on-site Hubbard interaction (bottom) for the conventional orbital embedding (a) and an unconventional embedding (b). In the latter, the geometry is no longer ideal, but the many-body spectrum is unchanged.}\label{fig_embedding}
\end{figure}

We now demonstrate the importance of orbital embedding on the ideal band property of the KM model.
For the KM model with flux $\phi = p/q < 1$, there are $q$ orbitals in each magnetic unit cell. The standard description of the model equally spaces these orbitals within the unit cell. However, one can adjust their relative locations while keeping the magnetic unit cell unchanged, which amounts to tuning the orbital embedding. Tuning the embedding is also equivalent to adjusting the Fourier transform convention for different orbitals \cite{jackson2015geometric}. 
As shown in Fig.~\ref{fig_embedding}, such a change ruins the ideal band property.
However, since the Hubbard interaction is purely on-site, tuning the relative locations of the orbitals does not affect the many-body ground state -- more concretely, the Laughlin state remains an exact zero-energy ground state after changing the embedding, despite violating the ideal band condition.
The many-body spectrum is also shown in Fig.~\ref{fig_embedding}.

In fact, changing the embedding transforms the flat band of the Kapit-Mueller model from an ideal band to a vortexable but non-ideal band with a nontrivial vortex function $z(\mathbf r)$.
This underscores the point that the geometric properties of bands cannot alone predict the FCI ground state~\cite{zhang2025beyond}.

\end{document}